\documentclass[11pt,a4paper]{article}
\usepackage{jheppub}
\pdfoutput=1
\usepackage{amsmath}
\usepackage{amsfonts}
\usepackage{amssymb}
\usepackage{mathtools}
\usepackage{graphicx}
\usepackage[utf8]{inputenc}
\usepackage{csquotes}
\usepackage{slashed}
\usepackage{booktabs}
\usepackage{tikz}
\usepackage[outline]{contour}
\contourlength{2.5pt}
\usepackage{xcolor}
\usepackage{siunitx}
\usepackage{hyperref}
\hypersetup{colorlinks, linkcolor = [rgb]{0,0.0,0.75}, citecolor = [rgb]{0,0.0,0.75}, urlcolor = [rgb]{0,0.0,0.75}}

\usetikzlibrary{decorations.markings}
\usetikzlibrary{arrows.meta,positioning}
\tikzset{%
->-/.style={decoration={markings,mark=at position #1 with {\arrow{>}}},postaction={decorate}},
->-/.default=0.5
}

\makeatletter
\g@addto@macro\bfseries{\boldmath}
\makeatother
\DeclareMathOperator{\Tr}{Tr}
\DeclareMathOperator{\tr}{tr}
\def\llangle{\langle\!\langle}
\def\rrangle{\rangle\!\rangle}
\def\obs{\mathcal O}
\def\dobs{\delta \obs}
\def\obsbar{\overline{\vphantom{(}\obs}}

\newcommand{\Cab}[2]{\llangle C_{#1} #2 \rrangle}
\renewcommand{\vec}{\boldsymbol}
\renewcommand{\tilde}{\widetilde}
\renewcommand{\Re}{\operatorname{Re}}
\newcommand{\dd}{\mathrm{d}}

\newcommand{\emiso}[0]{%
deDivitiis:2013xla,%
BMW:2014pzb,%
DiCarlo:2019thl,%
Borsanyi:2020mff,%
Boyle:2022lsi%
}

\newcommand{\spectral}[0]{%
Hansen:2017mnd,%
Hansen:2019idp,%
Bulava:2019kbi,%
Bruno:2020kyl,%
Bulava:2021fre,%
DelDebbio:2022qgu,%
Frezzotti:2023nun%
}

\AtBeginDocument{%
\heavyrulewidth=.08em
\lightrulewidth=.05em
\cmidrulewidth=.03em
\belowrulesep=.65ex
\belowbottomsep=0pt
\aboverulesep=.4ex
\abovetopsep=0pt
\cmidrulesep=\doublerulesep
\cmidrulekern=.5em
\defaultaddspace=.5em
}

\begin{document}

\title{
Exploiting stochastic locality in lattice QCD:\\ hadronic observables and their uncertainties
}

\author[a,b]{Mattia~Bruno,}
\affiliation[a]{Dipartimento di Fisica, Università di Milano-Bicocca, Piazza della Scienza 3, 20126 Milano, Italy}
\affiliation[b]{INFN, Sezione di Milano-Bicocca, Piazza della Scienza 3, 20126 Milano, Italy}

\author[a,b,c]{Marco~Cè,}
\affiliation[c]{Albert Einstein Center for Fundamental Physics (AEC) and Institut für Theoretische Physik, Universität Bern, Sidlerstrasse 5, 3012 Bern, Switzerland}

\author[d]{Anthony~Francis,}
\affiliation[d]{Institute of Physics, National Yang Ming Chiao Tung University, 30010 Hsinchu, Taiwan}

\author[e]{Patrick~Fritzsch,}
\affiliation[e]{School of Mathematics, Trinity College Dublin, Dublin 2, Ireland}

\author[f]{Jeremy~R.~Green,}
\affiliation[f]{Deutsches Elektronen-Synchrotron DESY, Platanenallee 6, 15738 Zeuthen, Germany}

\author[g]{Maxwell~T.~Hansen}
\affiliation[g]{School of Physics and Astronomy, University of Edinburgh, Edinburgh EH9 3FD, United Kingdom}

\author[h]{and Antonio~Rago}
\affiliation[h]{IMADA and Quantum Theory Center, University of Southern Denmark, Odense, Denmark}

\preprint{DESY-23-105}

\date{\today}

\abstract{
Because of the mass gap, lattice QCD simulations exhibit stochastic locality: distant regions of the lattice fluctuate independently. There is a long history of exploiting this to increase statistics by obtaining multiple spatially-separated samples from each gauge field; in the extreme case, we arrive at the master-field approach in which a single gauge field is used. Here we develop techniques for studying hadronic observables using position-space correlators, which are more localized, and compare with the standard time-momentum representation. We also adapt methods for estimating the variance of an observable from autocorrelated Monte Carlo samples to the case of correlated spatially-separated samples.
}

\makeatletter\gdef\@fpheader{~}\makeatother
\maketitle

\section{Introduction}

Numerical lattice QCD is, by now, a well established tool for extracting systematic and precise non-perturbative predictions of strong-force observables. The field has long sustained a positive feedback loop, with cutting edge calculations motivating continued technical advances, which in turn support the next generation of calculations.

The nature of the progress is varied; here we give three examples. First, the desire to push more observables into the sub-percent regime has led to recent developments in the use of multi-level algorithms \cite{Luscher:2001up,Ce:2016idq,Ce:2016ajy,Giusti:2017ksp}, which promise to exponentially improve the signal in importance-sampling-determined correlators, at fixed computational cost. A second example stems from the fact that many lattice calculations are already at percent or sub-percent precision: this has led to major developments in the inclusion of electromagnetic- and isospin-breaking effects, see e.g.~\cite{\emiso}, to ensure that meaningful quantities are being calculated at the reported precision. As a third example, going beyond the improvement of established lattice quantities, the field continues to develop strategies for new classes of observables. For example, the spectral methods described in refs.~\cite{\spectral} have received major attention recently as the basis of a new strategy for overcoming limitations of the Euclidean signature in calculations.

In this work we are concerned with another aspect of high-precision calculations: the need to design optimal estimators for both the central values and covariance matrices of lattice data. This is particularly relevant in the case of a limited number of gauge-field configurations.

The key concept that we exploit in this work is \emph{stochastic locality}, the notion that quantum fields belonging to regions sufficiently separated in space-time are exponentially decorrelated and fluctuate (almost) independently, thanks to the mass gap of QCD. In the limit of a single field configuration, generated with a finite but large volume capable to accommodate enough independent fluctuations of the fields, it is possible to exploit stochastic locality to define estimators of correlation functions and of variances from translation invariance \cite{Luscher:2017cjh}. This paradigm of performing so-called master-field simulations \cite{Luscher:2017cjh,Fritzsch:2021klm}, i.e.\ the generation of a few large-volume fields, is currently being investigated in the pure gauge theory \cite{Luscher:2017cjh,Giusti:2018cmp} and in full QCD~\cite{Fritzsch:2021klm,Fritzsch:2022lattice}. This idea was proposed in view of calculations at very fine lattice spacings, which suffer from frozen topological charge, leading to biases in the estimation of QCD observables. Working in a large volume suppresses this effect in addition to reducing systematic uncertainties from the periodicity of fields that occur independent of topological charge freezing.

The need to design optimal estimators can arise for multiple reasons. Such master-field calculations represent a fairly dramatic example, but in general, many modern lattice QCD calculations may face the practical need to rely on a small number of gauge-field configurations. For example, in practice, calculations using expensive actions such as domain-wall fermions (e.g.\ used extensively by the RBC-UKQCD and JLQCD collaborations), often end up generating fewer gauge-field configurations as compared to calculations with other actions.

This work represents a collection of methods and results that can be used to exploit their advantages in different lattice setups, i.e.\ their use is not limited to the master-field scenario. More specifically, we examine the following three aspects. First, we study how far stochastic locality can be pushed in traditional simulations with volumes between 4 and 9 in units of the pion Compton length, corresponding to 3 to 6 fm. Second, we examine strategies to design improved estimators of fermionic observables, with a specific attention to their volume-scaling properties, which are particularly relevant for the master-field program. Finally, we investigate whether typical low energy properties of simple hadrons, such as masses and transition matrix elements, can be extracted more efficiently from correlators defined in position space rather than the usual time-momentum representation (TMR).

This paper is organised as follows. In section~\ref{sec:basics} we present the details of the simulations used in this work and define the observables that we examine, both in time-momentum and coordinate-space representations. In section~\ref{sec:estimators} we introduce several estimators for such correlation functions (using stochastic locality as a guiding principle), review the formalism to calculate variances from spatial translation invariance and we present a detailed numerical study on its applicability. In section~\ref{sec:positionspace} we collect and work out the necessary formalism required to study correlators in position space, with a particular attention to boundary effects. For the extraction of the pion mass we show that the latter can be controlled analytically. In section~\ref{sec:results} we perform a detailed comparison of time-momentum representation and position-space data for the extraction of spectra and matrix elements. Finally, in section~\ref{sec:conclusions} we summarize and discuss our findings. The appendices contain several clarifications and collect additional results interesting for large volume simulations. Examples include truncated sums for momentum projections and an alternative implementation of the master-field idea using a large temporal extent.

\section{Lattice setup}
\label{sec:basics}

In this section we introduce the basic expectation values that we examine
in our study and define the notation that we will use throughout the manuscript. Further we describe the numerical setup used in the presented calculations.

\subsection{Hadronic correlation functions}
\label{sec:hadronic_correlators}

In the following we focus on simple pseudoscalar, axial, vector, and nucleon
two-point correlation functions in Euclidean space:
\begin{align}
C_{PP}(x) &\equiv \langle P(x) P^\dagger(0) \rangle, \label{eq:PPcorr} \\[2pt]
C_{AP,\mu}(x) &\equiv \langle A_\mu(x) P^\dagger(0) \rangle = \langle P(x) A_\mu^\dagger(0) \rangle ,\\[2pt]
C_{AA,\mu\nu}(x) &\equiv \langle A_\mu(x) A_\nu^\dagger(0) \rangle,\\[2pt]
C_{VV,\mu\nu}(x) &\equiv \langle V_\mu(x) V_\nu^\dagger(0) \rangle, \label{eq:VVcorr} \\[2pt]
C_{NN}(x) &\equiv \langle \chi(x) \bar\chi(0) \rangle, \label{eq:NNcorr}
\end{align}
where
\begin{align}
P(x) &\equiv \bar {u}(x) \gamma_5 d(x),\\
A_\mu(x) &\equiv \bar {u}(x)\gamma_\mu\gamma_5 d(x),\\
V_\mu(x) &\equiv \bar {u}(x) \gamma_\mu d(x),\\
\chi(x) &\equiv \epsilon_{abc} \big (u_a^T(x) C\gamma_5 d_b(x) \big ) u_c(x) \,.\label{eq:uud_spinor}
\end{align}
Here $u(x)$ and $d(x)$ are quark fields with Dirac and color indices left implicit, except for the color indices shown in the definition of $\chi(x)$. Both smeared and unsmeared quark fields are considered, as discussed further below.

The conventional approach in extracting physical information from such correlators is to use the time-momentum representation
of the two-point correlators:
\begin{equation}
\tilde C(t,\vec p) \equiv \int \dd^3\vec x \, e^{-i\vec p\cdot\vec x} \, C(\vec x,t).
\end{equation}
In our numerical studies, we will focus on $\vec p=\vec 0$
and simply denote that case
$\tilde C(t)\equiv \tilde C(t,\vec 0)$. Keeping generality for now, we are interested in the masses for the pseudoscalar and nucleon correlators obtained at large $t$:
\begin{align}
\tilde C_{PP}(t,\vec p) & \to \frac{|c_P(\vec p)|^2}{2E_\pi(\vec p)}
e^{-E_\pi(\vec p) t} \,, \\ %
\tilde C_{NN}(t,\vec p) & \to \frac{|c_N(\vec p)|^2}{2E_N(\vec p)}
(-i\slashed{p} + m_N)
e^{-E_N(\vec p)t} \,, %
\end{align}
where the $\rightarrow$ indicates that we only keep the leading term on the right-hand side. Here $p=(iE,\vec p)$ and $c_P$, $c_N$ are overlap factors that
parameterize the coupling of the interpolating operator to the ground
state.\footnote{If the quark field smearing is not $O(4)$ covariant,
two independent overlap factors are needed to describe the coupling
between the nucleon state and $\chi$~\cite{Bowler:1997ej,
Capitani:2015sba}. For analyzing time-momentum-representation
data, we will project $\tilde C_{NN}$ with $(1+\gamma_0)/2$, which
circumvents this problem.}
Up to exponentially suppressed volume effects, the masses can be extracted from $E_{\pi}(\boldsymbol p) = \sqrt{m_\pi^2 + \boldsymbol p^2}$ and $E_{N}(\boldsymbol p) = \sqrt{m_N^2 + \boldsymbol p^2}$.

From these correlation functions the pion decay constant $f_\pi$ is a further standard observable that can be determined. It is defined as the axial current matrix element with a properly normalised pion state, $\langle 0|A_\mu(0)|\pi(\vec p)\rangle\equiv ip_\mu f_\pi$.\footnote{%
We adopt the normalization convention in which the experimental physical value is $f_{\pi^\pm}^{\mathrm{exp}}\approx\SI{130}{\MeV}$.
}
If the quarks are local, i.e.\ not smeared, the $C_{AP,\mu}$ correlator can be used to extract $f_\pi$.
A conventional method to extract $f_\pi$ consists of taking the asymptotic part of the correlator in the time-momentum representation,
\begin{equation}
\tilde C_{AP,0}(t,\vec p) \to \frac{c_A(\vec p)c_P(\vec p)}{2E_\pi(\vec p)}
e^{-E_\pi(\vec p) t},
\end{equation}
with $\vec p = \vec 0$ and the axial current in the time direction $A_0$.
Then the decay constant is given by the ratio $c_A(\vec 0)/m_\pi$ up to renormalization and possible improvement.
Since in this paper we are interested only in computing and comparing the noise of the matrix element, here we neglect such complications and simply define
\begin{equation}
f_\pi^{\mathrm{bare}} = \frac{c_A}{m_\pi} \,,
\end{equation}
where the $c_A=c_A(\vec 0)$ amplitude (as well as $c_P=c_P(\vec 0)$) can be chosen to be real and positive.
At the practical level, one can compute $f_\pi^{\mathrm{bare}}$ from a combined fit of $\tilde C_{PP}(t,\vec 0)$ and $\tilde C_{AP,0}(t,\vec 0)$ with $m_\pi$, $c_A$ and $c_P$ as parameters.

For the vector correlator, we decompose
the spatial components at zero momentum as
$\tilde C_{VV,ij}(t,\vec 0)\equiv \delta_{ij}\tilde C_{VV}(t)$; if the quark fields
are unsmeared, this can be used to obtain the isovector part of the
subtracted hadronic vacuum polarization (HVP) function \cite{Bernecker:2011gh}:
\begin{equation}
\label{eq:subtracted_HVP_TMR}
\bar \Pi(-Q^2) = \int_0^\infty \dd t
\left[ t^2 - \frac{4}{Q^2}\sin^2\left(\frac{Qt}{2}\right) \right]
Z_V^2 \tilde C_{VV}(t),
\end{equation}
where $Z_V$ is the renormalization factor for the vector current. Also in this case we neglect possible improvement terms.

\subsection{Numerical setup}

To verify our analytical findings and to effectively study the
presented methods, we generate a number of ensembles that enable
like-by-like comparisons. These ensembles span a range of volumes at
fixed pion mass and a range of pion masses at fixed volume so that the
corresponding scaling behaviors can be studied in detail.

All ensembles are generated using the stabilised Wilson fermion framework \cite{Francis:2019muy}, which in particular employs the exponentiated clover action for the fermions.
For the gauge fields we adopt the L\"uscher-Weisz action \cite{Weisz:1982zw,Luscher:1984xn,Curci:1983an}.
Throughout, we use $N_f=2+1$ dynamical fermion flavors at the single lattice spacing of $a \simeq 0.094~\mathrm{fm}$ ($\beta=6/g_0^2=3.8$) \cite{Francis:2019muy}. The lattice spacing is determined via the gradient-flow scale $t_0$ and converted to physical units via pseudoscalar decay constants \cite{Bruno:2016plf}.
The clover coefficient $c_{SW}(g_0^2)$ is tuned non-perturbatively \cite{Francis:2019muy}.

The ensembles that we used or generated for this study are reported in table~\ref{tab:lat_ens}
and span pion masses ranging from (approximately) $410~\mathrm{MeV}$ down to $215~\mathrm{MeV}$ and volumes from
$m_\pi L \simeq 3.3$ up to 9.
The largest spatial extent we reach has a physical length $L\simeq 6~\mathrm{fm}$.
For recent progress in generating larger lattices
with physical volumes of $L\simeq 9~\mathrm{fm}$ and $L\simeq 18~\mathrm{fm}$,
following the master-field paradigm, see refs.~\cite{Fritzsch:2021klm,Fritzsch:2022lattice}.

The chiral trajectory is set by fixing the sum of quark masses at the so-called flavor symmetric point ~\cite{Bietenholz:2010jr, Bruno:2014jqa,Strassberger:2021tsu}.
This implies that the strange quark mass is lighter than the physical value. We do not consider observables containing an explicit strange quark here and focus on purely light-quark (isovector) correlators.
We further draw attention to the 32BT ensemble: here the same space-time volume of the 64B ensemble is reached
as from our smallest volume by elongating the $T$ direction while keeping the spatial volume fixed at $L=32a$. We discuss results on this long-$T$ ensemble in appendix \ref{sec:app_longT}; see also ref.~\cite{Bruno:2022ljo} as initial reference for the long-$T$ approach.

We note that quark field smearing is a technique commonly used when studying hadrons. It serves two purposes: First and foremost, it suppresses contributions from excited states relative to the ground state, so that the latter can be isolated at shorter distances. Second, it extends the footprint of point-source propagators, which tends to improve the statistical signal. The standard smearing methods used with the time-momentum representation extend only in spatial directions. As such they are $O(3)$-covariant but not $O(4)$-covariant. To study how these common methods behave in our approach, we adopt the smearing from ref.~\cite{Papinutto:2018ajw}, which uses a fermion propagator in three dimensions. In order to study hadrons in position space, we also use quark fields with the gradient flow~\cite{Luscher:2013cpa} applied, which is an $O(4)$-covariant smearing.
Details on smearing parameters are given below in table~\ref{tab:point_sources_results_1} and the surrounding text in section~\ref{sec:results_position_space}.

\begin{table}[tb]
\centering
\begin{tabular}{lccccccccc}
\toprule
Label & $\beta$ & $L/a$ & $T/a$ & $\kappa_u$ & $\kappa_s$ & $m_\pi[\rm{MeV}]$ & $m_\pi L$ & $V/V_0$ & $N_\mathrm{MC}$\\\hline
32A & 3.8 & 32 & 96 & 0.1389630 & 0.1389630 & 410 & 6.3 & 1.0 & 100 \\
32B & & 32 & 96 & 0.1391874 & 0.1385164 & 293 & 4.5 & 1.0 & 100 \\
32C & & 32 & 96 & 0.1392888 & 0.1383160 & 215 & 3.3 & 1.0 & 100 \\
\midrule
32BT& & 32 & 768 & 0.1391874 & 0.1385164 & 293 & 4.5 & 8.0 & 50 \\
\midrule
48B & & 48 & 96 & 0.1391874 & 0.1385164 & 293 & 6.7 & 3.4 & 100 \\
64B & & 64 & 96 & 0.1391874 & 0.1385164 & 293 & 8.9 & 8.0 & 50 \\
\bottomrule
\end{tabular}
\caption{$N_f=2+1$ gauge ensembles used in this work. All simulations use the stabilised Wilson fermion framework. The coefficient of the exponentiated clover is set to $c_{SW}=1.955242$. $V/V_0$ denotes the ratio of the global lattice volume w.r.t.\ to $32^3 \times 96$. For ensemble 64B we discuss in appendix~\ref{app:exceptional} the identification of an exceptional configuration earlier than these 50.}\label{tab:lat_ens}
\end{table}

\section{Estimators of correlators and variances}
\label{sec:estimators}

In this section we continue with the description of the
numerical strategies adopted to estimate both central values and errors,
while numerical results are deferred to the next sections.

In Lattice QCD calculations, Wick's theorem is employed to express
fermionic correlation functions in terms of quark propagators $S(x,y)$. Point-source propagators are defined from sources with support
on a single lattice site, and may be used to calculate two-point
isovector correlators on a fixed gauge-field background as
\begin{equation}
\mathcal C^\mathrm{point}(x;y) \equiv - \Re \Tr\big[\Gamma' S(x+y,y) \Gamma S(y,x+y) \big].
\end{equation}
We omit from the notation spin, color and
flavor indices, since we are only interested in observables depending on the two light degenerate
quark fields, and we also do not indicate whether the underlying quark fields have been smeared with a specific label, but specify it
in the text.

Taking the expectation value\footnote{
Depending on the context, we use $\langle \dots \rangle$ to denote
either a QCD expectation value or a Monte Carlo expectation
value. The latter runs over gauge fields $U$ and also possibly over
any noise fields used for stochastically estimating fermionic observables.}
over fluctuations of the gauge field, we obtain
\begin{equation}
C(x) = \langle \, \mathcal C^\mathrm{point}(x;y) \, \rangle
\end{equation}
where $C$ may be taken from eqs.~(\ref{eq:PPcorr}--\ref{eq:VVcorr})
depending on the choice of $\Gamma'$ and $\Gamma$.
To improve the signal of such expectation
values, one often makes use of translation invariance to displace
the location $y$ of the source and to average the corresponding different
estimators of the correlation function.
Stochastic locality suggests that to obtain efficient sampling with largely
uncorrelated samples,
the minimal distance among the source locations should be a function
of the typical correlation length of the system, and of the observable.
Taken together with translation invariance, one naturally arrives at a regular
displacement of the point sources on a grid, which we discuss below.

\begin{figure}[t]
\centering
\begin{tikzpicture}[scale=0.8]
\clip (-1.5,-0.5) rectangle (16.5,6.5);
\draw[thin,dotted] (-5,-5) grid (20,10);
\draw[dashed,step=5,xshift=2.5cm,yshift=2.5cm] (-5,-5) grid (20,10);
\draw[thin,dashed] (5,5) circle [radius=2.5];
\draw[thin,<->] (-0.5,0) -- (-0.5,5) node[midway] {\contour{white}{$b$}};
\draw[thin,<->] (5,5) -- (7.0,3.5) node[midway,right] {\contour{white}{$r_{\text{max}}=b/2$}};
\draw[thick,->-] ( 5,5) to [bend right] ( 4,3);
\draw[thick,->-] ( 4,3) to [bend right] ( 5,5);
\draw[thick,->-,lightgray] ( 5,0) to [bend right=25] ( 4,3);
\draw[thick,->-,lightgray] ( 4,3) to [bend right=25] ( 5,0);
\draw[thick,->-,lightgray] (10,0) to [bend right=15] ( 4,3);
\draw[thick,->-,lightgray] ( 4,3) to [bend right=15] (10,0);
\foreach \x in {0,5,10,15}
\foreach \y in {0,5}
\fill (\x,\y) circle [radius=3pt];
\draw (5,5) node[above] {\contour{white}{$y\in G$}};
\draw[fill=white] ( 4,3) circle [radius=3pt] node[below left] {\contour{white}{$x$}};
\end{tikzpicture}
\caption{Sketch of the estimator $\mathcal{C}^\text{sgrid}$ with a grid of point sources over a two-dimensional window of the lattice. The set $G$ of source points $\protect\tikz{\fill (0,0) circle(2.4pt);}\in G$ is a regular grid with spacing $b$. A mesonic two-point function is evaluated at sink point $x$ that is in the domain defined by $y\in G$ and within a distance $r_{\text{max}}$ from $y$. Two of the spurious contributions from the \enquote{wrong} source are shown in light gray.}\label{fig:sources_sketch}
\end{figure}
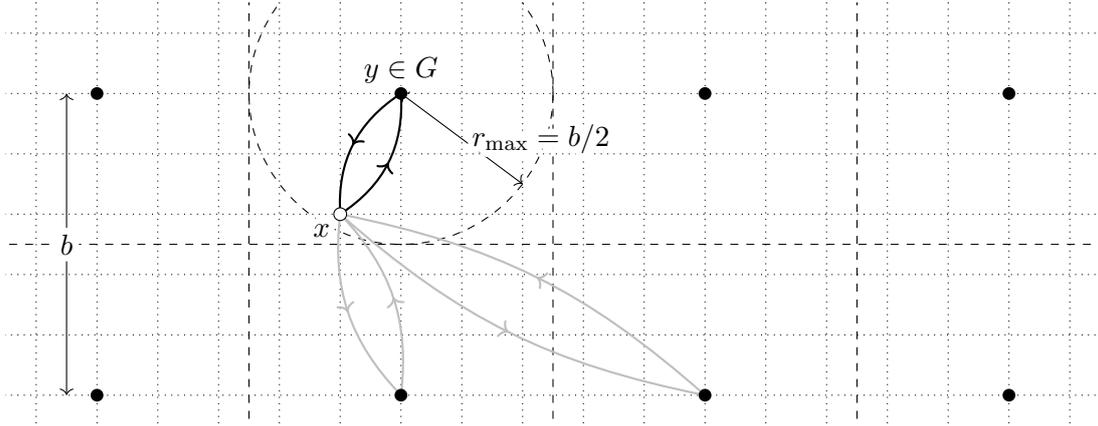

\subsection{Stochastic estimators}
\label{sec:stochastic_estimators}

Solving for propagators from many different point sources $y\in G$ has
a cost that scales proportionally to $|G|$, the number of such
points. To avoid such cost scaling, one could let the source have
support on all points in $G$ and perform a single propagator solve.
For a given sink location $x$, the correlator would be dominated by the closest
source $y$, and in general the contributions from the other
points of the grid would be exponentially suppressed: see figure~\ref{fig:sources_sketch}.
However, when $x-y$ grows, disentangling their effect from the main
signal would become increasingly difficult;
therefore, it is desirable to explicitly eliminate such contaminations.
To do so, we introduce $N_\eta$ noise fields $\eta_i(x)$ with support on
the sparse set of points $G$ that define a \emph{stochastic grid}, which satisfy
\begin{equation}
\frac{1}{N_\eta} \sum_i \eta_i(x) \eta_i^\dagger(y) =
\langle \eta_i(x) \eta_i^\dagger(y) \rangle + O(N_\eta^{-1/2}) \,,
\end{equation}
with
\begin{equation}
\langle \eta_i(x) \rangle = 0 \,, \quad
\langle \eta_i(x) \eta_j^\dagger(y) \rangle = \delta_{ij} \delta_{xy} I_{sc}
\quad \text{for } y \in G,
\label{eq:eta}
\end{equation}
where $I_{sc}=I_s\otimes I_c$ is the identity matrix in spin and
color space. The neglected terms of $O(N_\eta^{-1/2})$ represent
the stochastic noise.
Letting $\psi_i(x)$ be the propagator with $\eta_i$ as its source, we obtain
\begin{equation}
\langle \psi_i(x) \eta_j^\dagger(y) \rangle = \delta_{ij} S(x,y) \,,
\quad \text{for } y \in G\,.
\end{equation}
In practice, we employ U(1) noise with color and spin
dilution~\cite{Wilcox:1999ab, Foley:2005ac}, which can be represented
as having $\eta_i(x)$ be a diagonal spin-color matrix containing
random phases.
In this work we consider the estimator
\begin{equation}
\label{eq:grid_estimator}
\mathcal{C}^\text{sgrid}(x;y) \equiv
\frac{-1}{N_\eta(N_\eta-1)} \sum_{i\neq j} \Re \Tr
\left[\psi_i^\dagger(x+y) \gamma_5 \Gamma' \psi_j(x+y) \eta_j^\dagger(y) \Gamma \gamma_5 \eta_i(y) \right] \,,
\end{equation}
which satisfies $\langle \mathcal C^\text{sgrid}(x;y) \rangle = C(x)$
for all $y \in G$. Therefore, on a single configuration the correlator averaged from all $|G|$
points in $G$,
\begin{equation}
\llangle \mathcal C^\text{sgrid}(x) \rrangle = \frac{1}{|G|} \sum_{y \in G}
\mathcal C^\mathrm{sgrid} (x;y) \,,
\label{eq:C_estimator}
\end{equation}
defines an estimator for $C(x)$ with stochastic error that scales as
$|G|^{-1/2}\propto V^{-1/2}$ if the volume is increased while keeping
the grid spacing constant. While having improved cost scaling, it
differs from the average over point source estimators for all $y\in G$
by an additional $O(N_\eta^{-1/2})$ stochastic error. Here we have
used $\llangle \cdot \rrangle$ to denote a volume-averaged estimator;
this notation will be especially used in the next subsection.

For a baryon, the correlator has the form,
$C_{NN}(x-y)=\langle B[S(x,y),S(x,y),S(x,y)]\rangle$, where $B$ is a
trilinear map containing the color and spin contractions,
which may be implicitly defined from
eqs.~\eqref{eq:NNcorr} and~\eqref{eq:uud_spinor}.
Similarly to eq.~\eqref{eq:grid_estimator} a stochastic estimator is defined from
\begin{equation}
\begin{split}
\mathcal C_{NN}^\text{sgrid}(x-y;y) = & \frac{1}{N_\eta(N_\eta-1)(N_\eta-2)}
\\ & \times \sum_{i,j,k\text{ distinct}} B[
\psi_i(x)\eta_i^\dagger(y),\psi_j(x)\eta_j^\dagger(y),\psi_k(x)\eta_k^\dagger(y)] \,.
\end{split}
\end{equation}
For our choice of color- and spin-diluted $U(1)$ noise, the diagonal
nature of $\eta_i(y)$ for each $y\in G$ allows us to use the simpler
(and still unbiased) estimator
\begin{equation}
\mathcal{C}_{NN}^{\text{sgrid}'}(x-y;y)=B[S_\text{avg}(x,y),S_\text{avg}(x,y)S_\text{avg}(x,y)]\,,
\end{equation}
with
\begin{equation}
S_\text{avg}(x,y) = \frac{1}{N_\eta} \sum_i \psi_i(x)\eta_i^\dagger(y), \qquad y\in G \,.
\end{equation}

The equations derived above work for any distribution of the
source locations and (depending on the specific problem) different displacements
from a regular grid might turn out to be more efficient.
The (potential) advantage
of the estimator in eq.~\eqref{eq:grid_estimator}
is that fewer than $|G|$ explicit evaluations of propagators may be sufficient.
For short-distance observables, with a small footprint, denser grids
lead to a substantial reduction of the variance (see for instance ref.~\cite{Blum:2015you}).
On the other hand, for long-distance observables sparser grids may be preferable; in
this case, field configurations with large volumes profit much more from
such a strategy. Specifically, if one keeps the density of points in the grid constant,
the cost grows only linearly in the volume.

In practice, with master fields
the calculation might be significantly
accelerated by considering a block decomposition of the Dirac operator\footnote{
If properly tuned it may lead to significant speed-up factors by reducing
communications. Similar strategies are currently being explored in the context
of the HMC algorithm in ref.~\cite{Boyle:2022ncb}.
},
along the lines of refs.~\cite{Ce:2016idq,Giusti:2022xdh},
for example by centering the domain around the location
of a point of the grid~\cite{Luscher:2017cjh}.
We do not investigate this approach here and defer its study to future
work.

The estimators defined above are designed to obtain an optimized
sampling for correlators that depend generically on the four-vector $x$.
However physical information is commonly extracted from the
time-momentum representation. In this case it is more efficient to
let the noise field have support on all points at fixed Euclidean time,
which we call a \emph{stochastic wall},
and to use the one-end trick~\cite{Foster:1998vw} to obtain an estimator
that requires just one noise field.
In this case, we let $\eta_i(x)$ be a scalar noise field with support on the
wall and $\psi_i(x)$ be the matrix field containing the propagator with
$\eta_i\otimes I_{sc}$ as its source~\cite{ETM:2008zte}.
Denoting the Euclidean time location of the wall with $y_0$,
a simple estimator for the isovector
zero-momentum correlator of two bilinear operators
is found as:
\begin{equation}
\begin{split}
\widetilde {\mathcal C}^\mathrm{swall}(x_0-y_0; \vec x) = &
-\frac{1}{N_\eta} \sum_{i}
\Re \Tr \big[\Gamma \gamma_5 \psi_i(x)^\dagger \gamma_5 \Gamma' \psi_i(x) \big] \\ = &
- a^3 \sum_{\vec y}
\Re \Tr \big[\Gamma S(y,x) \Gamma' S(x,y) \big]
+ O(N_\eta^{-1/2}) \,.
\label{eq:swall}
\end{split}
\end{equation}
Since $\eta_i(x)$ has support on an entire time slice,
zero-momentum
projection is achieved at the source, and $\widetilde {\mathcal C}^\mathrm{swall}(x_0-y_0; \vec x)$ satisfies
$\langle \widetilde {\mathcal C}^\mathrm{swall}(x_0-y_0; \vec x) \rangle = \widetilde C(x_0-y_0)$
for all $\vec x$.
Therefore the freedom at the sink location, i.e.\ $\vec x$,
can be used to define the estimator
\begin{equation}
\llangle \widetilde{\mathcal C}^\mathrm{swall}(x_0-y_0) \rrangle =
\frac{a^3}{L^3} \sum_{\vec x} \mathcal C^\mathrm{swall}(x_0-y_0; \vec x) \,,
\end{equation}
which after the gauge-field average has an improved variance
compared to the same time-momentum correlator calculated from point sources,
for fixed moderate computational cost.
The fine sampling of $(L/a)^3$ points will be particularly useful for
a detailed study in section~\ref{sec:masterfield_error_saturation} of
the saturation of our estimate of the variance of
$\llangle\widetilde{\mathcal C}^\text{swall}\rrangle$.

\subsection{Variances}
\label{sec:variances}

The above definitions of the correlation functions induce a situation where the observables
(indexed by $\alpha$, $\beta$, \dots) have localized estimators\footnote{
The index $\alpha$ can include information such as the separation $y$ of
operators in a two-point correlation function, for the case where the
observable $\obs_\alpha(x)$ is the estimator $\mathcal{C}(y;x)$.
}
$\obs_\alpha(x)$, and data are available for $N = |\Lambda|$ regularly-spaced
points $x \in \Lambda$. Here $N$ could be as large as $(L/a)^4$ if all lattice
points are available, but our analysis is generic and also applies to
situations where $\Lambda$ is a $D$-dimensional subspace or a
coarse sub-grid of spacing $b$, in which case $N=(L/b)^D$. As such it covers the three mentioned implementations using points, grids and walls.

In our case, we restrict ourselves to the
situation where $\obs_\alpha(x)$ are known on a single large-volume field configuration.
The best estimator of the true expectation value $\langle \obs_\alpha \rangle$
is given by the volume average
\begin{equation}
\llangle \obs_\alpha \rrangle \equiv \frac{1}{N} \sum_{x} \obs_\alpha(x) ,
\label{eq:obs_alpha}
\end{equation}
where the sum is to be understood as running over the $N$ points
in $\Lambda$ for which data are available.
Using invariance under space-time translations, the covariance
of our volume-average estimators \cite{Luscher:2017cjh},
\begin{equation}
\langle
\big[ \llangle \obs_\alpha \rrangle - \langle \obs_\alpha \rangle \big]
\big[ \llangle \obs_\beta \rrangle - \langle \obs_\beta \rangle \big]
\rangle = \frac{1}{N^2} \sum_{x,y} \Gamma_{\alpha\beta}(x-y) =
\frac{1}{N} \sum_y \Gamma_{\alpha\beta}(y) \, ,
\label{eq:Gamma}
\end{equation}
is given in terms of the correlation function $\Gamma_{\alpha\beta}$,
\begin{equation}
\Gamma_{\alpha \beta}(y) \equiv
\langle \big[\obs_\alpha(y) - \langle \obs_\alpha \rangle \big]
\big[\obs_\beta(0) - \langle \obs_\beta \rangle \big] \rangle\,,
\end{equation}
in analogy to the autocorrelation function for Monte Carlo time
\cite{Wolff:2003sm}.
Likewise, we define $C_{\alpha\beta}$ as the sum of
$\Gamma_{\alpha\beta}(y)$ without the factor of $1/N$,
\begin{equation}
C_{\alpha\beta} \equiv \sum_y \Gamma_{\alpha\beta}(y)
= \sum_{y \in \text{inf.\ vol.}} \Gamma_{\alpha\beta}(y) \Bigl(1 + O(e^{-m_\pi L}) \Bigr),
\end{equation}
where the finite-volume sum can be approximated as an infinite-volume
sum, up to exponentially suppressed corrections.
The function $\Gamma_{\alpha\beta}$ is expected to fall off exponentially as a function
of the distance $|y|$, with a mass $m$ that is dictated by the details of the system. In particular, because $\mathcal O_\alpha$ has a non-zero vacuum expectation value, states with vacuum quantum numbers will appear in the spectrum and the asymptotic behavior will typically be determined by the lightest $0^{++}$ state,%
\footnote{We thank R.~Sommer for pointing this out.}
i.e.\ $m=2m_\pi$. %
In principle, this approach leads to improved error estimators, i.e.\ a reduction of
the error of the error, compared to the traditional Monte Carlo
analysis where the space-time information is blocked and not exploited in the
error estimates.

In a practical situation, where the volume is large but finite, we have to
substitute $\langle \obs_\alpha \rangle$ with
$\llangle \obs_\alpha \rrangle$, thus obtaining a (biased)
estimator for $\Gamma_{\alpha \beta}$ given by
\begin{equation}
\llangle \Gamma_{\alpha\beta}(y) \rrangle \equiv \frac{1}{N} \sum_{x}
\dobs_\alpha(x+y) \, \dobs_\beta(x) \,, \quad
\dobs_\alpha(x) \equiv \obs_\alpha(x) - \llangle \obs_\alpha \rrangle \,,
\label{eq:Gavg}
\end{equation}
with bias (see appendix \ref{app:errors})\footnote{The generalization of eq.~\eqref{eq:Gavg} to cases where the observables
are known only on an irregular subset $\Lambda$ of points is
\begin{equation}
\llangle\Gamma_{\alpha\beta}(y)\rrangle \equiv
\frac{ \sum_{x,x' \in \Lambda}
\dobs_\alpha(x) \, \dobs_\beta(x') \, \delta_{x+y,x'} }{
\sum_{x,x' \in \Lambda} \delta_{x+y,x'} }.
\end{equation}}
\begin{equation}
\big\langle \llangle \Gamma_{\alpha\beta}(y) \rrangle \big\rangle - \Gamma_{\alpha\beta}(y) =
- \frac{C_{\alpha\beta}}{N} \,.
\end{equation}
Moreover,
it is necessary to truncate the sum defining the covariance matrix:
we introduce the
finite summation radius $R$ and define
\begin{equation}
\Cab{\alpha\beta}{(R)}
\equiv \sum_{|y|\leq R} \llangle \Gamma_{\alpha\beta}(y) \rrangle.
\label{eq:Cbar}
\end{equation}
The truncation introduces an additional bias:
\begin{equation}
\bigl\langle \Cab{\alpha\beta}{(R)} \bigr\rangle
= C_{\alpha\beta} \Bigl( 1 + O(e^{-mR}) -N(R)/N \Bigr),
\end{equation}
where $m$ is the mass governing the falloff of $\Gamma_{\alpha\beta}(y)$.
Where appropriate (such as for $\alpha=\beta$), in analogy to the integrated
autocorrelation time $\tau_\text{int}$~\cite{Wolff:2003sm},
we also introduce the integrated correlation volume
\begin{equation}
\tau_\alpha(R) \equiv \sum_{|y|\leq R} \frac{\Gamma_{\alpha\alpha}(y)}{\Gamma_{\alpha\alpha}(0)}
\,,
\label{eq:tauD}
\end{equation}
such that the variance of $\llangle \obs_\alpha \rrangle$ becomes
\begin{equation}
\operatorname{var}( \llangle \obs_\alpha \rrangle ) \approx
\frac{\tau_\alpha(R)}{N} \Gamma_{\alpha\alpha}(0) \,,
\label{eq:var_obs}
\end{equation}
with generically $\operatorname{var}(X)\equiv\langle (X-\langle X\rangle)^2\rangle$.
Note that $\tau_\alpha(R)$
implicitly depends on the number of dimensions $D$ of the subspace where the
master-field analysis is considered, and for $D=1$, this has the same form
as $\tau_\mathrm{int}$.

Extending the derivation of ref.~\cite{Wolff:2003sm} to $D$ dimensions
(see appendix \ref{app:errors}),
we obtain for the error of
the error
\begin{equation}
\operatorname{var}(\Cab{\alpha\beta}{(R)}) \approx
\frac{N(R)}{N} \big[
C_{\alpha\alpha}C_{\beta\beta} +
C_{\alpha\beta}^2 \big] \,.
\label{eq:Cbar_err}
\end{equation}
Above $N(R)$ is the number of available points satisfying $|y|\leq R$ and
can be approximated as $V_D(R/b)$ where %
$V_D(r)={\pi^{D/2} r^D}/{\Gamma(D/2+1)}$ is the volume of a $D$-ball of radius $r$ and $\Gamma(x)$ is Euler's gamma function.
This result can be used to formulate an automatic windowing procedure
for a master-field type analysis where an optimal summation radius $R$
is found by balancing statistical and systematic errors; see appendix~\ref{app:window}.
In summary: for the chosen $R$, our estimate for the covariance of
$\llangle\obs_\alpha\rrangle$ and $\llangle\obs_\beta\rrangle$ is
$\Cab{\alpha\beta}{(R)}/N$, which for $\alpha=\beta$ is also given
by eq.~\eqref{eq:var_obs}.

Finally to conclude our considerations on master-field error estimators,
we note that
a \emph{blocking} procedure can be defined in analogy to binning
of adjacent gauge configurations in a Monte Carlo chain.
That is, one may use a method to absorb the effect of autocorrelations so to treat the blocked data with standard statistical tools.
Specifically starting from a $N=(L/a)^D$ lattice, a coarse blocked lattice
of size $N_B=(L/b)^D$, is obtained from
\begin{equation}
\obs_{B\alpha}(u) \equiv \frac{N_B}{N} \sum_{x \in \mathrm{block}\, u} \obs_\alpha(x) \,,
\end{equation}
with $u_\mu$ the coordinate of the block. At this point the analysis
can proceed as before with the replacement $x_\mu \to u_\mu$ and
$N \to N_B$. As expected, by increasing the block size
$\Cab{\alpha\beta}{(R)}$ saturates at smaller $N(R)$: this corresponds
to smaller $R/b$, but the physical scale $R$ governing saturation
remains the same. Thus $\llangle\Gamma_{\alpha\beta}(0)\rrangle$ becomes an
increasingly better approximation of the error, at the expense of a
larger error of the error (see also ref.~\cite{Wolff:2003sm}). For
sufficiently large block sizes, the blocks become statistically
independent and standard analysis techniques such as bootstrap or
jackknife can be used without the $D$-dimensional formalism.
On the other hand, the downside of blocking is that the error-of-the-error due to correlations is suppressed by a power law rather than exponentially.
In this
study we used minimal blocking, i.e.\ $b/a=2$ or $4$, only as a
practical strategy to reduce storage costs for the observables and to
speed up intermediate stages of the analysis; this enables the
calculation of $\llangle\Gamma_{\alpha\beta}\rrangle$ without the need for a
computing cluster.

\subsection{Error saturation in master-field estimates}
\label{sec:masterfield_error_saturation}

To study the scaling of the summation radius $R$ with our ensembles
with traditional volumes and several field configurations,
we estimate the modified correlation function\footnote{
Using instead $\delta \obs_\alpha^i(x) = \obs_\alpha^i(x) -\llangle \obs_\alpha^i \rrangle$ would introduce a large bias in our current smaller-than-required volumes. With $\llangle \obsbar_\alpha \rrangle$ we further suppress
it by a factor of $1/N_\mathrm{MC}$, improving the convergence of our plateaus.
}
\begin{equation}
\llangle \Gamma^i_{\alpha\beta}(y) \rrangle = \frac{1}{N}
\sum_{x \in \Lambda} \delta \obs_\alpha^i(x) \delta \obs_\beta^i(x+y) \,,
\quad
\delta \obs^i_\alpha(x) = \obs^i_\alpha(x) - \llangle \obsbar_\alpha \rrangle
\label{eq:Gbar}
\end{equation}
on every Monte Carlo configuration $i$, where
\begin{equation}
\llangle \obsbar_\alpha \rrangle \equiv \frac{1}{N_\mathrm{MC} N} \sum_i \sum_{x \in \Lambda} \obs^i_\alpha(x) \,.
\end{equation}
Here $\Lambda$ denotes the subset of points considered in the specific analysis
and $N_\mathrm{MC}$ the total number of independent measurements.
When autocorrelations are sizeable, the index $i$ in eq.~(\ref{eq:Gbar})
runs over bins of sufficiently long length and $N_\mathrm{MC}$ is redefined as the number
of bins.
Exploiting the additional dimension offered by Monte Carlo time,
we estimate the error of $\Gamma_{\alpha \beta}$ itself using the variance of
$\llangle \Gamma_{\alpha \beta}^i \rrangle$ among the $N_\text{MC}$ available (decorrelated)
samples in Monte Carlo time, which returns more precise estimates than the
analytic formula in eq.~(\ref{eq:Cbar_err}). In master-field calculations,
the latter may be more useful.

\begin{figure}[t]
\centering
\includegraphics{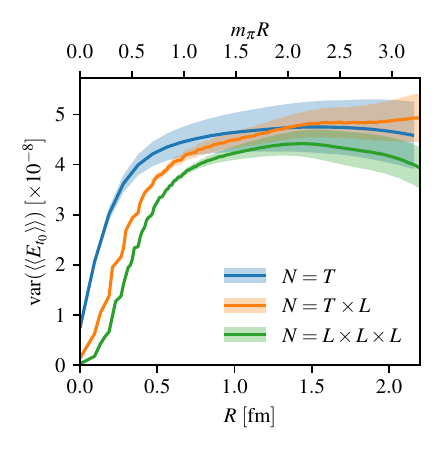}
\includegraphics{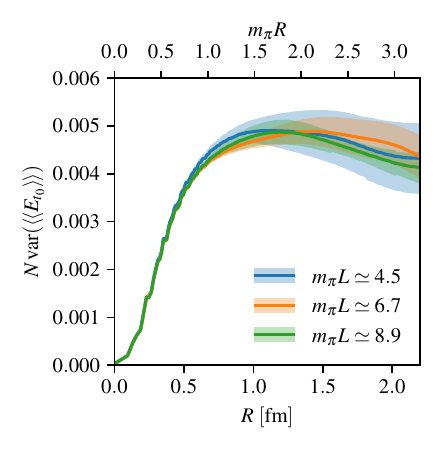}
\caption{Variance of the estimator $\llangle E_t \rrangle $ for a flow time $t\approx t_0$. \textbf{Left}: variance estimated using stochastic locality on our 48B ensemble for different choices of $\Lambda$ given in eq.~\eqref{eq:var_T_TL_LLL}. \textbf{Right}: scaling of the variance with the volume, estimated on the 32B, 48B and 64B ensembles using $\Lambda_{L^3}$.}\label{fig:Et}
\end{figure}

The first observable that we examine is the gradient-flowed energy density
$E_t$ at positive flow time $t\simeq t_0$~\cite{Luscher:2010iy}.
We consider the symmetric definition based on the clover discretization
of the field strength tensor.
Being a pure-gauge one-point function, it can be evaluated for all space-time
points, thus giving us the opportunity to study the
saturation of the variance with $R$ for different estimators based on
different $D$-dimensional partitions $\Lambda$ of the lattice.
Specifically, we examine the three cases
\begin{equation}
\begin{array}{ll}
\Lambda_T &= \{ x_0 | x_0 \in [0,T-a] \} \,, \\
\Lambda_{TL} & = \{ (x_0,x_1) | x_0 \in [0,T-a], x_1 \in [0,L-a] \} \,, \\
\Lambda_{L^3} & = \{ \vec x | x_1,x_2,x_3 \in [0,L-a] \} \,, \\
\end{array}
\label{eq:var_T_TL_LLL}
\end{equation}
and we block the data on the corresponding orthogonal dimensions:
for example, in the case of $\Lambda_T$,
$E_t(x)$ is pre-averaged over the three spatial directions.
In the left panel of figure~\ref{fig:Et}, we show how these lower dimensional
partitions of the lattice can be used to estimate the
error based on stochastic locality\footnote{Variances and errors of dimensionful quantities are plotted in lattice units.}. We observe, when we increase the
number of dimensions
the variance starts to saturate at larger values of $R$.
This effect is understood as a consequence of the $D$-dimensional integration measure $\dd^D x \propto \dd R \, R^{D-1}$.
The presence of clear plateaus suggests
that error estimators based on stochastic locality
are already accessible (for some quantities) in several present calculations with
moderate volumes, for example for correlation functions known on several or all source time locations.

For the energy density this statement is further supported by the right
panel of figure~\ref{fig:Et} where we compare our three volumes
and observe excellent agreement of the plateaus obtained for the variance.
In fact, even though the study of $E_{t_0}$ in a real master-field pure-gauge simulation
was already presented in ref.~\cite{Luscher:2017cjh},
here we numerically demonstrate, in dynamical QCD, that for our precision the plateau is reached at $m_\pi R \simeq 2$ implying that
volumes as small as $m_\pi L \simeq 4$ are sufficiently large
to saturate its master-field error.

\begin{figure}[t]
\centering
\includegraphics{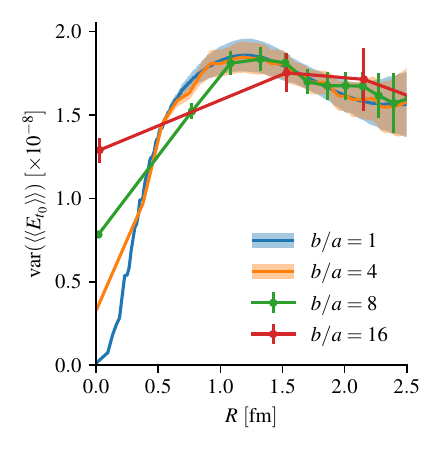}%
\includegraphics{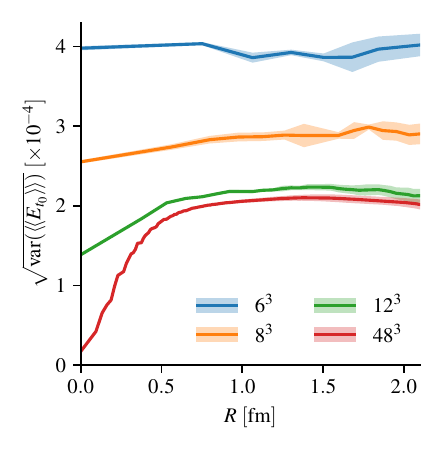}
\caption{Variance of $\llangle E_{t_0} \rrangle$ calculated from a three-dimensional (spatial) slice, i.e.\ using $\Lambda_{L^3}$ in eq.~\eqref{eq:var_T_TL_LLL}. \textbf{Left}: scaling of the variance with the blocking size $b/a$ for the 64B ensemble. The case $b/a=1$ amounts to considering all $(L/a)^3$ points. \textbf{Right}: scaling of the error with the sparsity of the three-dimensional grid. The red curve corresponds to taking all $N=48^3$ points of a time slice of the 48B ensemble.}\label{fig:Et_blocking}
\end{figure}

In figure~\ref{fig:Et_blocking}, left panel, we demonstrate how the blocking
procedure outlined in section~\ref{sec:variances} effectively resums spatial correlations,
leading to a faster fall-off of $\Gamma_{\alpha \beta}$.
In addition, the error of the error at the saturation $R$ is comparable between different block sizes and no blocking, except for the largest $b=16a$, where we observe a larger error of the error.
This makes modest blocking a cost-efficient solution to storage and memory constraints. At the same time, the fact that the error is less well determined for the largest blocking indicates that error estimates based on stochastic locality should provide an improvement as compared to estimates based on fluctuations across distinct gauge fields.

On the fermionic side, we will be considering simple mesonic two-point correlators
projected to zero momentum, with pseudoscalar and
vector operators, cf.~eqs.~\eqref{eq:PPcorr} and \eqref{eq:VVcorr}.
One strategy to estimate the corresponding correlation function $\Gamma_{\alpha\beta}(x)$
could be in principle based on several point sources located on a regular grid, as
discussed in section~\ref{sec:stochastic_estimators}. Taking into account the cost to calculate
quark propagators, interesting questions to answer are how many point sources should be considered,
and at which point should we stop. In the right panel of figure~\ref{fig:Et_blocking} we examine
this scenario using the energy density as our probe.
Despite knowing it for all locations,
we perform a sub-sampling over three-dimensional regular grids of various sizes to imitate the
case of an estimator based on stochastic grids (in the limit of
large $N_\eta$).
As expected, by making the grid denser, we reduce the variance and eventually
start to sample the short-distance structure of $\Gamma_{\alpha\beta}(x)$. This is similar to traditional Monte Carlo averages where we resolve autocorrelations by measuring more frequently in Molecular-Dynamics units.
At this point we saturate the amount of independent statistical information
and can stop increasing the number of point sources.
For this example, the error based on $N=12^3$ samples is nearly as small as that based taking all $N=48^3$ points while using 64 times fewer samples.

\begin{figure}[t]
\centering
\includegraphics{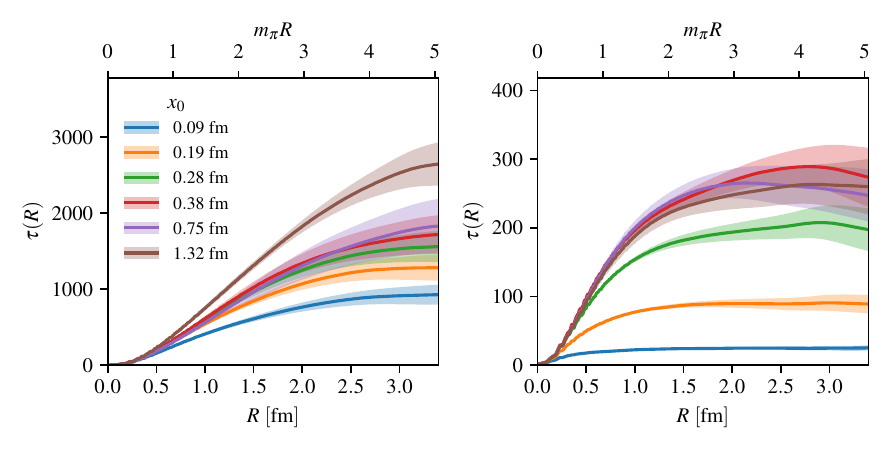}
\caption{Integrated correlation volume $\tau(R)$ from eq.~\eqref{eq:tauD} as a function of the source-sink separation $x_0$ and summation radius $R$, for zero-momentum projected correlators measured on the 64B ensemble using stochastic walls. \textbf{Left}: pseudoscalar two-point function, $\widetilde C_{PP}(x_0)$. \textbf{Right}: vector-vector correlator $\widetilde C_{VV}(x_0)$.}\label{fig:cvar}
\end{figure}

To study the saturation window from fermionic observables, we find it more
practical to adopt standard approaches based on stochastic wall sources.
From the identification $\obs_{\alpha = x_0}(\vec x)$
with $\widetilde{\mathcal C}^\mathrm{swall} (x_0;\vec x)$ taken from eq.~\eqref{eq:swall}, at
fixed $x_0$ we use the sink location $\vec x$
to estimate the master-field error in an $L^3$ subvolume [i.e.\ $\Lambda_{L^3}$ in eq.~\eqref{eq:var_T_TL_LLL}].

The two-point function with zero-momentum projection at the source is a non-local
observable. Heuristically we expect that, as one increases the source-sink separation, the ``footprint'' of this quantity in the finite spacetime volume grows.
To illustrate this, figure~\ref{fig:cvar} shows the integrated
correlation volume introduced in eq.~\eqref{eq:tauD} for the
pseudoscalar and vector-vector correlators and a range of source-sink
separations $x_0$.
In general we observe a growth of $\tau$ with larger values of $x_0$, and
correspondingly a larger summation radius needed to reach the plateau in $\tau(R)$.
For the vector-vector correlator, the expected scaling of the
correlation function $\Gamma(x)$ is with $e^{-2m_\pi |x|}$.
We checked that our data are compatible with such scaling by
fitting $\tau(R)$ with the fit function $c \Gamma(s,M R)$, with $c,s,M$ as free
parameters, and $\Gamma(s,b)$ defined in eq.~\eqref{eq:gammainc}.
Note that the apparent saturation in $x_0$ reached for $x_0 \geq 0.38 ~\mathrm{fm}$
should not be taken as a fact: this observable is
affected by the signal-to-noise problem which means that at fixed statistics
(and volume) for sufficiently large $x_0$ the noise will be large enough
to hide the effect of increasingly larger correlation volumes,
in the same manner as autocorrelations along Monte Carlo time emerge
from less noisy observables.

The pseudoscalar correlator, on the contrary, does not suffer from the signal-to-noise
problem and is much more likely to show a poor convergence in $R$ within our volumes.
We demonstrate this in the left panel of figure~\ref{fig:cvar}.
By increasing the source-sink separation the footprint of the observable grows up
to the point where very short plateaus are observed, eventually preventing the
usage of this error estimation technique for the study of the asymptotic behavior
of the correlator.
This suggests that additional statistical
information is injected only by new field configurations or larger volumes, i.e.\ master fields.

\begin{figure}[t]
\centering
\includegraphics{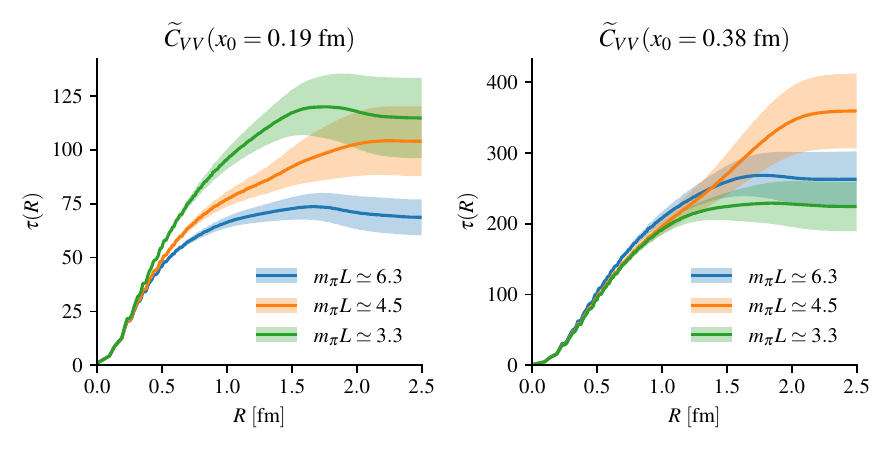}
\caption{Integrated correlation volume $\tau(R)$, cf. eq.~\eqref{eq:tauD}, of the vector-vector correlator at short (left panel) and long distances (right panel), as a function of the pion mass and summation radius $R$. The three ensembles 32A, 32B, and 32C have the same volume but differ in the simulated values of the quark masses, with $m_\pi L$ ranging from 3.3 up to 6.3.}\label{fig:tau_mpi}
\end{figure}

The pion mass plays an important role in the applicability of this error estimation technique since it governs the exponential convergence of $\tau(R)$ at large values of $R$.
By lowering the light quark masses we expect a larger footprint of the master-field variance estimator, or in other words, a larger $\tau(R)$. This phenomenon is observed in the left panel of figure~\ref{fig:tau_mpi}, where we plot the scaling with the pion mass of the integrated correlation volume $\tau(R)$ for the vector-vector two-point function $\tilde C_{VV}(x_0)$ at short source-sink separation, using our three $32^3$ lattices, with $m_\pi L \simeq 3.3, 4.5$ and 6.3, cf. table~\ref{tab:lat_ens}. As expected, by decreasing the pion mass, plateaus are reached for increasingly larger values of $R$. The same behavior is apparently not observed in the right panel, where we examine the correlator at a longer source-sink separation. To understand why this is the case, we must first consider the fact that at longer distances and fixed statistics, we expect larger statistical errors as we approach the chiral limit. With noisier data, resolving spatial correlations at longer distances becomes more challenging and as a consequence our three (asymptotic) estimators of $\tau$ become compatible. We conclude that the effect seen in the right panel of figure~\ref{fig:tau_mpi} is only an artefact of the lower relative statistical accuracy on lighter ensembles and does not imply a different hierarchy.

We conclude this study of the variance based on stochastic locality with
a few considerations.
First we note that it is a powerful tool
also for traditional volumes with $m_\pi L \simeq 4$. Specifically for simulations
with a low number of (independent) gauge field configurations, reliable
error estimators may be obtained from master-field
analysis exploiting invariance under one or more directions, depending on the
data available. Clearly one has to examine every observable (e.g.\ every
source-sink separation) independently and check that long reliable
plateaus can be identified. First explorations along these lines can be found
in ref.~\cite{RBC:2023pvn}.
Since the complexity of this strategy
quickly grows with the number of observables, in appendix~\ref{app:window}
we discuss an extension of the automatic window procedure proposed
by Wolff~\cite{Wolff:2003sm} to this context.

Second, the correlation function $\Gamma_{\alpha\beta}(x)$ is an observable of the theory with a well-defined infinite-volume limit and exponentially suppressed volume corrections.
As such, its $x$ dependence may be studied (at fixed lattice spacing)
on intermediate volumes and used in the planning of the number of
fields and volumes required for a given master-field simulation
and precision.
{For example, by fitting the variance of the pion correlator on the 64B ensemble for $x_0=4a\simeq0.38~\mathrm{fm}$,
we can approximately estimate its asymptotic value.
From the latter we deduce that a master-field analysis
on a three-dimensional space with geometry $(L/a)^3 = 192^3$
would return a pion correlator with similar half-percent accuracy
from a single configuration.}

\section{Position-space correlators}
\label{sec:positionspace}

It is reasonable to expect that a local-in-space-time approach to correlators might be beneficial in exploiting the stochastic locality properties of fields to the fullest extent possible. Here, we consider fully position-space correlation functions as one possible strategy to achieve this. By not including a momentum projection,
we avoid explicitly introducing long-distance contributions to the
correlators. If the quark field smearing is $O(4)$
covariant, then the overlap factors are independent of $\vec
p$. Neglecting discretization and finite-volume effects, this allows
us to take the inverse of the three-dimensional Fourier transform and
obtain for $|x|\to\infty$:
\begin{align}
\label{eq:CPPdef}
C_{PP}(x) &\rightarrow \frac{c_P^2}{4\pi^2} \frac{m_\pi}{|x|} K_1(m_\pi|x|),\\
\label{eq:CAPdef}
C_{AP,\mu}(x) &\rightarrow \frac{c_A c_P}{4\pi^2} \frac{x_\mu}{|x|} \frac{m_\pi}{|x|} K_2(m_\pi|x|),\\
\label{eq:CAAdef}
C_{AA,\mu\nu}(x) &\rightarrow \frac{c_A^2}{4\pi^2} \left[ -\delta_{\mu\nu} \frac{1}{x^2} K_2(m_\pi|x|) + \frac{x_\mu x_\nu}{x^2} \left( \frac{m_\pi}{|x|} K_1(m_\pi|x|) + \frac{4}{x^2} K_2(m_\pi|x|) \right) \right]\,,\\
\label{eq:CNNdef}
C_{NN}(x) &\rightarrow \frac{c_N^2}{4\pi^2} \frac{m_N^2}{|x|}\left[
K_1(m_N|x|) + \frac{\slashed{x}}{|x|} K_2(m_N|x|) \right] .
\end{align}
Above, $K_n(x)$ denotes a modified Bessel function of the second kind.
We make use of the (Euclidean) Lorentz invariance of the theory in the continuum\footnote{
Here we assume continuum physics in a very large volume.
The boundary effects are discussed later in this section, while the practical implementation of the angular averaging at finite lattice spacing is introduced in section~\ref{sec:results_position_space}.
} to introduce correlators that are functions of the $4d$ radial direction $r\equiv |x|$ only and transform as scalars.
For each of $C_{PP}(x)$ and $C_{AP,\mu}(x)$ there is only one option, and we use a \,$\mathring{}$\, label to indicate the corresponding function of $r$
\begin{gather}
\mathring{C}_{PP}(r) = C_{PP}(x)
\,, \label{eq:CPPring_def} \qquad
\mathring{C}_{AP}(r) = x_\mu C_{AP,\mu}(x) \,.
\end{gather}
The asymptotic behavior for $r \to \infty$ is then
\begin{gather}
\mathring{C}_{PP}(r)
\rightarrow \frac{c_P^2}{4\pi^2} \frac{m_\pi}{r} K_1(m_\pi r)\,, \label{eq:Cring_asym} \qquad
\mathring{C}_{AP}(r)
\rightarrow \frac{c_A c_P}{4\pi^2} m_\pi K_2(m_\pi r) \,.
\end{gather}

For $C_{AA,\mu\nu}(x)$, by contrast, there are two possible contractions, leading to
\begin{align}
\mathring{C}_{AA}^{(1)}(r) & = \delta_{\mu\nu} C_{AA,\mu\nu}(x)
\ \rightarrow \ \frac{c_A^2}{4\pi^2} \frac{m_\pi}{r} K_1(m_\pi r) , \label{eq:CAAring1_def} \\
\mathring{C}_{AA}^{(2)}(r) & = x_\mu x_\nu C_{AA,\mu\nu}(x)
\ \rightarrow \ \frac{c_A^2}{4\pi^2} \left[ m_\pi r K_1(m_\pi r) + 3 K_2(m_\pi r) \right] . \label{eq:CAAring2_def}
\end{align}
Similarly, the spinor-valued correlator $C_{NN}(x)$ can give rise to two contractions
\begin{align}
\mathring{C}_{NN}^{(1)}(r) & = \tr C_{NN}(x)
\ \rightarrow \ \frac{c_N^2}{4\pi^2} \frac{m_N^2}{r} K_1(m_N r) , \label{eq:CNNring1_def} \\
\mathring{C}_{NN}^{(2)}(r) & = \tr\slashed{x}C_{NN}(x)
\ \rightarrow \ \frac{c_N^2}{4\pi^2} m_N^2 K_2(m_N r) . \label{eq:CNNring2_def}
\end{align}

Finally, it was shown in ref.~\cite{Meyer:2017hjv} how to determine
the ($I=1$ component of the) subtracted HVP function $\bar\Pi(-Q^2)=\Pi(-Q^2)-\Pi(0)$ at space-like momenta $Q^2>0$ from the (light quarks) vector correlator in position space:
\begin{multline}
\label{eq:PiHVP_CCS}
\bar\Pi(-Q^2) = \int \dd^4x
\left[ x^2\delta_{\mu\nu} f_1(Q^2|x|^2) - x_\mu x_\nu f_2(Q^2|x|^2) \right]
Z_V^2 C_{\mu\nu}(x) \\
= 4\pi^2 Z_V^2 \int_0^\infty r^3 \dd r \left[ r^2 f_1(Q^2r^2) \mathring{C}_{VV}^{(1)}(r) - f_2(Q^2r^2) \mathring{C}_{VV}^{(2)}(r) \right] ,
\end{multline}
for known kernels $f_{1,2}$ and
\begin{gather}
\mathring{C}_{VV}^{(1)}(r) = \delta_{\mu\nu} C_{\mu\nu}(x) \,, \label{eq:CVVring_def} \qquad
\mathring{C}_{VV}^{(2)}(r) = x_\mu x_\nu C_{\mu\nu}(x) .
\end{gather}

At this stage it is instructive to incorporate the
leading wrap around finite-volume effects that occur as $r$ approaches
$L/2$ (while continuing to neglect discretization effects). For the pseudoscalar correlator in the time-momentum
representation, the leading finite-temporal-extent effect can be
accounted for by modifying the effective mass to include the expected
$\cosh$-like behavior. A similar approach can be applied to
position-space correlators that have a pseudoscalar as the leading
long-distance contribution; we will see that this involves integrals
of Bessel functions.

In the case of the pseudoscalar correlator, the leading position-space boundary effects are given by summing over images of the scalar propagator $C_{PP}(x)$, defined in eq.~\eqref{eq:CPPdef}, in the four periodic directions. To make this concrete we define
\begin{equation}
C_{PP}^{\mathbb L}(x) \equiv \sum_{n} C_{PP}(x + \mathbb L \cdot n) \,,
\end{equation}
where $\mathbb L$ is a diagonal matrix encoding the lattice geometry and $n\in\mathbb{Z}^4$ is a four-vector of integers.
The introduction of a finite periodic space-time breaks continuous rotational symmetry, so $C_{PP}^{\mathbb L}(x)$ is not automatically a function of $r\equiv |x|$ as it is in eq.~\eqref{eq:CPPring_def}.
One possible strategy to address this is to explicitly perform an average over points with fixed $r$.
To this end we introduce
\begin{equation}
\mathring{C}_{PP}^{\mathbb L}(r) = \frac{1}{2 \pi^2} \int \dd\Omega_4 \, C_{PP}^{\mathbb L}(x) = \frac{1}{2 \pi^2} \sum_n \int \dd\Omega_4 \, C_{PP}(x + \mathbb L \cdot n) \,,
\end{equation}
where $\int\dd\Omega_4=\int_0^\pi\dd\varphi_1\sin^2\varphi_1\int_0^\pi\dd\varphi_2\sin\varphi_2\int_0^{2\pi}\dd\varphi_3$.
To simplify, one can perform a change of variables for each fixed $n$ such that $n$ is aligned with the $z$-axis. Evaluating the $\dd\varphi_2$ and $\dd\varphi_3$ integrals, we reach a one-dimensional integral in $\varphi\equiv\varphi_1$,
\begin{equation}
\int \dd\Omega_4 \, C_{PP}(x + \mathbb L \cdot n) = 4\pi\int_0^\pi \dd \varphi \sin^2 \varphi \cdot \mathring C_{PP}(|x^{\mathbb L}_{n}|(r,\varphi)) \,,
\end{equation}
where we have used that the infinite-volume correlator $C_{PP}(x)$ is in fact only a function of $\vert x \vert$, and have also made use of $\mathring C_{PP}(r)$ as defined in eq.~\eqref{eq:CPPring_def}. We have additionally introduced the shorthand
\begin{equation}
\label{eq:xLn_def}
|x^{\mathbb L}_{n}|(r,\varphi) = |x + \mathbb L \cdot n| = \sqrt{r^2+(\mathbb L \cdot n)^2+2|\mathbb L \cdot n|r\cos\varphi} \,.
\end{equation}
It follows that the angular averaged finite-volume correlator, $\mathring{C}_{PP}^{\mathbb L}(r)$, is equal to
\begin{equation}
\mathring{C}_{PP}^{\mathbb L}(r) = \frac{2}{\pi} \sum_n \int_0^\pi \dd \varphi \sin^2 \varphi \cdot \mathring C_{PP}(|x^{\mathbb L}_{n}|(r,\varphi)) .
\end{equation}

Taking now the specific case of a $L^3 \times T$ geometry, we drop terms scaling as $e^{- \sqrt{2 }m_{\pi} L}$, $e^{- m_\pi T}$ or faster and substitute the asymptotic form for $ \mathring C_{PP}(r )$, eq.~\eqref{eq:CPPdef}, to reach a concrete approximation for $ \mathring{C}_{PP}^{\mathbb L}(r)$. At large distances $0 \ll r \leq L/2$, this takes the form $\vert c_P \vert^2 F_{PP}(r, m_\pi, L)$, where
\begin{equation}
F_{PP}(r, m_\pi, L) = \frac{1}{4\pi^2} \frac{m_\pi}{r} K_1(m_\pi r)
+ 6 \, \frac{2}{ \pi } \int_0^\pi \dd \varphi \sin^2 \varphi \frac{1}{4\pi^2} \frac{m_\pi}{ |x^{L}_{\hat z}|(r,\varphi)} K_1(m_\pi |x^{L}_{\hat z}|(r,\varphi)) %
\,,
\label{eq:FPP_final}
\end{equation}
with
\begin{equation}
\label{eq:xLz_def}
|x^{L}_{\hat z}|(r,\varphi) = \sqrt{r^2+ L^2 + 2L r \cos\varphi} \,,
\end{equation}
defining a special case of eq.~\eqref{eq:xLn_def}.

This concrete functional form can be used to define a position-space analogue of the cosh effective mass, by numerically determining the unique function $M$ satisfying
\begin{equation}
\label{eq:numerical_effective_mass}
M\bigg ( \frac{ F_{PP}(r, m_\pi, L) }{ F_{PP}(r + \Delta, m_\pi, L) }, r, \Delta, L \bigg ) = m_\pi \,.
\end{equation}
In figure~\ref{fig:pion_effmass}, below, we show the numerical results of implementing this strategy; see also section~\ref{sec:position_space_results_pion}.

In the case of the $C_{AP,\mu}(x)$ correlator, one can follow a similar procedure.
The radial correlator in infinite volume is defined in eq.~\eqref{eq:CPPring_def}, with a factor of $x_\mu$ to contract the $\mu$ index of the axial current $A_\mu$.
The corresponding finite-volume expression summed over images is
\begin{equation}
\mathring C_{AP}^{\mathbb L}(r) = \frac{1}{2 \pi^2} \sum_n \int \dd\Omega_4 \, x_\mu C_{AP,\mu}(x + \mathbb L \cdot n) \, ,
\end{equation}
where it is important to note that the contracted $ x_\mu$ is introduced by hand and is therefore not summed over the periodic images.

Following exactly the same recipe as above, one finds that the asymptotic behavior of $\mathring C_{AP}^{\mathbb L}(r) $ on a $L^3 \times T$ geometry can be written as $c_A c_P F_{AP}(r, m_\pi, L)$ where
\begin{equation}
F_{AP}(r, m_\pi, L) = \frac{ m_\pi}{4\pi^2} K_2(m_\pi |x^{L}_{\hat z}|(r,\varphi)) + 6 \frac{2}{\pi} \int_0^\pi \dd \varphi \sin^2 \varphi \ f^{L}_{\hat z}(r,\varphi) \
\frac{ m_\pi}{4\pi^2} K_2(m_\pi |x^{L}_{\hat z}|(r,\varphi)) \,.
\end{equation}
In addition to $|x^{L}_{\hat z}|$, defined in eq.~\eqref{eq:xLz_def} above, this estimator depends on
\begin{gather}
f^{L}_{\hat z}(r,\varphi) = \frac{r^2+ L r\cos\varphi}{r^2+L^2+2 L r \cos\varphi} \,,
\end{gather}
which is a special case of $f^{\mathbb L}_{n}(r,\varphi) = {x \cdot (x + \mathbb L \cdot n)}/{(x + \mathbb L \cdot n)^2}$ .

The final step is to perform the same derivation for $C_{AA, \mu \nu}(x)$. In eqs.~\eqref{eq:CAAring1_def} and~\eqref{eq:CAAring2_def} above we have defined the functions $\mathring{C}_{AA}^{(1)}(r)$ and $\mathring{C}_{AA}^{(2)}(r)$, resulting from two scalar contractions of the indices $\mu$ and $\nu$, together with angular averaging. The finite-volume quantities with the same definitions are then given by
\begin{align}
\mathring C_{AA}^{\mathbb L(1)}(r) & = \frac{1}{2 \pi^2} \sum_n \int \dd\Omega_4 \, \delta_{\mu\nu} C_{AA,\mu\nu}(x + \mathbb L \cdot n) \,, \\
\mathring C_{AA}^{\mathbb L(2)}(r) & = \frac{1}{2 \pi^2} \sum_n \int \dd\Omega_4 \, x_\mu x_\nu C_{AA,\mu\nu}(x + \mathbb L \cdot n) \,.
\end{align}

The asymptotic volume effects for each finite-volume correlator $\mathring C_{AA}^{\mathbb L(i)}(r)$ can be expressed in terms of a known function multiplying the axial-current overlap factor. This function is given by
$\vert c_A \vert^2 F_{AA}^{(i)}(r, m_\pi, L)$.
For the case of $i=1$, in which the indices are contracted, one finds the same result as for the pseudoscalar (PP) correlator, $F^{(1)}_{AA}(r, m_\pi, L) = F_{PP}(r, m_\pi, L) $. By contrast, for the case of $i=2$, the result is more complicated
\begin{align}
F^{(2)}_{AA}(r, m_\pi, L) & = \frac{1}{4\pi^2} \left[ m_\pi r K_1(m_\pi r) + 3 K_2(m_\pi r) \right] \nonumber \\[5pt]
& + 6 \,\frac{2}{\pi} \int_0^\pi \dd \varphi \sin^2 \varphi \ [f^{L}_{\hat z}(r,\varphi)]^2 \ \frac{1}{4\pi^2} \left[ m_\pi \zeta K_1(m_\pi \zeta) + 3 K_2(m_\pi \zeta) \right]_{\zeta = |x^{L}_{\hat z}|(r,\varphi)} \nonumber
\\[5pt]
& + 6 \, \frac{2}{\pi} \int_0^\pi \dd \varphi \sin^2 \varphi \ g^{L}_{\hat z}(r,\varphi) \ \frac{1}{4\pi^2} 3 K_2(m_\pi |x^{L}_{\hat z}|(r,\varphi)) \,,
\end{align}
where
\begin{equation}
g^{L}_{\hat z}(r,\varphi) = -\frac{1}{3} \frac{L^2r^2\sin^2\varphi}{(r^2+ L^2+2 L r\cos\varphi)^2} \,,
\end{equation}
is a special case of
\begin{equation}
g^{\mathbb L}_{n}(r,\varphi) = \frac{1}{3} \frac{[x_\mu(x + \mathbb L \cdot n)_\mu]^2-x^2(x + \mathbb L \cdot n)^2}{(x + \mathbb L \cdot n)^4} \,. %
\end{equation}

To reveal the exponential scaling of these effects, we expand the special functions introduced above about $L = \infty$, and obtain for eq.~\eqref{eq:FPP_final}
\begin{equation}
F_{PP}(r, m_\pi, L) = \frac{1}{4\pi^2} \frac{m_\pi}{ r} \bigg [ \sqrt{\frac{\pi }{2 m_\pi r}} e^{-m_\pi r}
+ 6 \frac{ e^{- m_\pi (L-r)}}{ (m_\pi r)^{1/2} (m_\pi L)^{3/2}} \bigg ]
\,,
\end{equation}
where we have expanded using $L \gg r \gg 1/m_\pi$ and have dropped all subdominant terms.
This implies that the volume corrections can be significantly enhanced if $r$ is taken too large at fixed $L$. For typical volume sizes of $m_\pi L \simeq 4$, such effects must be included in the analysis. For the ensembles used in this work, the effect is still quite significant (as shown in figure~\ref{fig:pion_effmass}, below) and is larger than in the case of zero-momentum projection, for which the backward-propagating pion is suppressed. The latter is likely due to the fact that the temporal extent is larger than the spatial extent on all ensembles included.
Nonetheless, the importance of wrap around effects in the position-space correlator highlights how such methods benefit significantly more (from the point of view of systematic errors) from larger volumes.

The corresponding corrections for the nucleon mass are significantly more complicated. The dominant volume effects arise from the nucleon emitting and re-absorbing a pion, not from the mirror images, and therefore depend on the nucleon-pion-nucleon coupling. Quantifying these effects goes beyond the scope of the present work and this issue arises in general for all states that couple to pions, e.g.\ for multi-pion states in $C_{PP}(x)$.

\section{Numerical results}
\label{sec:results}

In the first subsection we investigate the stochastic grids as a potential efficient estimator of position-space correlators on large volumes.
In the second subsection we instead examine differences between position-space correlators and more traditional momentum-projected ones computed on the same point sources for several standard mesonic and baryonic observables.

\subsection{Stochastic grids}
\label{sec:efficient_estimators}

Let us divide the lattice into equal domains centered around the
regularly displaced source points of our grid.
The stochastic estimator in eq.~\eqref{eq:grid_estimator} is in principle valid for any
sink point $x$, but is particularly efficient for source-sink separations that do not exceed
such domains.
To test the efficiency of the stochastic grids introduced above,
we use our larger ensemble with lattice volume of $96\times 64^3$ sites,
and define $G$ as a grid of $3\times 2^3=24$ points with spacing $b=32a$.
The $|G|$ domains centered around each $y\in G$ are hypercubes of $32^4$ sites and
we study several radial correlators for distances with norm $r \leq 16 a$.

\begin{figure}[t]
\centering
\includegraphics{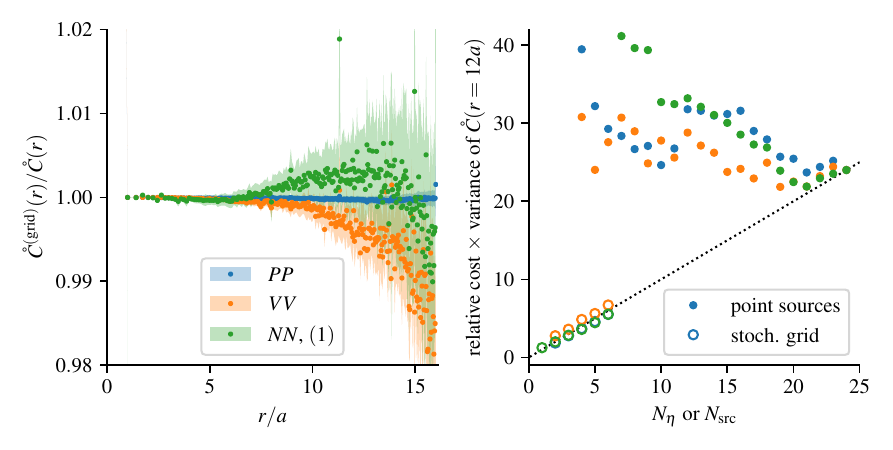}
\caption{In this figure we compare different estimators for several correlators, cf. eqs.~(\ref{eq:CPPring_def}),~(\ref{eq:CNNring1_def}) and~(\ref{eq:CVVring_def}), defined in position space as a function of the radial distance. Specifically we consider up to 6 random sources with support on a $3 \times 2^3$ grid, with fixed offset, and up to 24 point sources gradually placed to fill the same point locations of the grid. \textbf{Left}: ratio of the estimator based on $N_\eta=6$ stochastic grids over the estimator based on 24 point sources. Correlated errors are represented by the error bands. \textbf{Right}: product of computational cost and variance for the stochastic grid estimator (empty markers) as a function of $N_\eta$ compared to the point sources estimator (solid markers) as a function of $N_{\mathrm{src}}$. For each choice of pseudoscalar, vector or nucleon correlator (in the same colors as in the plot on the left), the cost $\times$ variance product is relative to $1/24$th of the cost $\times$ variance of $N_{\mathrm{src}}=24$ point sources.}\label{fig:point_grid}
\end{figure}

In the left panel of figure~\ref{fig:point_grid}, we study the efficiency of the
stochastic grid estimator by comparing it with the average of up to 24
point-source estimators using exactly the points in $G$ and the same
gauge field configurations. The leading contribution to the
stochastically vanishing difference between the two estimators will
come from the neighboring points in $G$. For this comparison, we
examine the pseudoscalar, vector, and nucleon radial correlators. In
the left panel, we use the most precise estimator in each case (6
stochastic grid sources or 24 point sources) and show the ratio of the
two estimators as a function of the radial distance. Within its
uncertainty (and within \SI{\pm 2}{\percent}), it is consistent with
one as should be the case for our unbiased estimators. The noisy
deviation from one is very small for small $r$ and grows (particularly
for the nucleon and vector correlators) at larger $r$; this is to be
expected, since as $r$ increases, the correlator decays while the
sink becomes closer to the neighboring sources responsible for this
deviation.

In the right panel of figure~\ref{fig:point_grid}, we compare the product of computational cost and variance of the grid estimator of $\mathring C(r)$ at a fixed $r=12a$ with the one with the point estimator, for the three correlators $\mathring C_{PP}$, $\mathring C_{VV}$ and $\mathring C^{(1)}_{NN}$.
This is plotted as a function of the number of either noise sources $N_\eta$ or point sources $N_{\mathrm{src}}$ for the two estimators respectively. Both are equal to the number of 12-component solutions of the Dirac equation.
As expected, the product of cost and variance is approximately constant with increasing $N_{\mathrm{src}}$ for the point estimator.
Note that they have been arbitrarily normalized to 24 units at $N_{\mathrm{src}}=24$.
In the same units, the product of cost and variance of the the grid estimator is significantly lower and grows approximately linearly with $N_\eta$.
At $N_\eta=6$ the variance of the grid estimator is between 3 and 5.5 times smaller than the one of the point estimator at the same cost, with the gain being larger for the vector and nucleon correlators than for the pion one.
On larger volumes we expect stochastic grids to be even more efficient, which makes them a promising strategy for master-field calculations.

Even though the volumes studied in this work are fairly large, if we employ stochastic grids with our minimal setup $L/b=2$, we are only able to examine position-space correlators up to radial distances of 16 lattice units. For this reason, we turn to
estimators based on single point sources to test position-space methods
against correlators in the time-momentum representation.
This allows us to instead reach $r \leq 32a$, a distance that is sufficient to reliably extract --- once boundary effects are taken into account --- ground-state energies like the pion and nucleon masses, as shown in sections~\ref{sec:position_space_results_pion} and~\ref{sec:nucleon_mass} respectively.
Further tests of stochastic grids on much larger volumes up to
$m_\pi L \simeq 25$ are presented in ref.~\cite{Ce:2023sqk}.

\subsection{Position-space and time-momentum representations}
\label{sec:results_position_space}

\begin{table}[tb]
\centering
\begin{tabular}{llS[table-format=1.3]S[table-format=1.5(2)]S[table-format=1.3(2)]S[table-format=1.3(1)]}
\toprule
& smearing & {$t/a^2$} & {$am_\pi$} & \multicolumn{2}{c}{$am_N$} \\
& & {or $\kappa_{\mathrm{3d}}$} & & {$\tr C_{NN}$} & {$\tr\slashed{x}C_{NN}$} \\
\midrule
position-space & & & 0.13764(38) & 0.486(11) & 0.508(10) \\
position-space & grad. flow & 1.227 & 0.13754(35) & 0.494( 6) & 0.506( 5) \\
\midrule
zero-momentum & & & 0.13807(38) & 0.64 ( 6) & \\
zero-momentum & 3d ferm. & 0.180 & 0.13819(34) & 0.51 ( 4) & \\
zero-momentum & 3d ferm. & 0.190 & 0.13817(35) & 0.50 ( 4) & \\
zero-momentum & 3d ferm. & 0.200 & 0.13812(35) & 0.46 ( 4) & \\
\bottomrule
\end{tabular}
\caption{Pion and nucleon masses from position-space correlators on point sources, compared to zero-momentum projected correlators on the same point sources, with or without smearing.}\label{tab:point_sources_results_1}
\end{table}

Below we present results obtained from the 64B ensemble. Its spatial volume in units of the inverse pion mass is
approximately nine, and from our previous studies on the saturation of
the error we already know that it is not sufficient for the pion
correlator. Therefore, in the following we calculate errors along
Monte Carlo time, using traditional analysis strategies.
We compare results from position space versus TMR. We also show
the effect of smearing, using $3d$ fermions~\cite{Papinutto:2018ajw} for TMR and gradient flow
for position space; parameters are given in
table~\ref{tab:point_sources_results_1}.

\subsubsection{Pion mass}
\label{sec:position_space_results_pion}

\begin{figure}[t]
\centering
\includegraphics{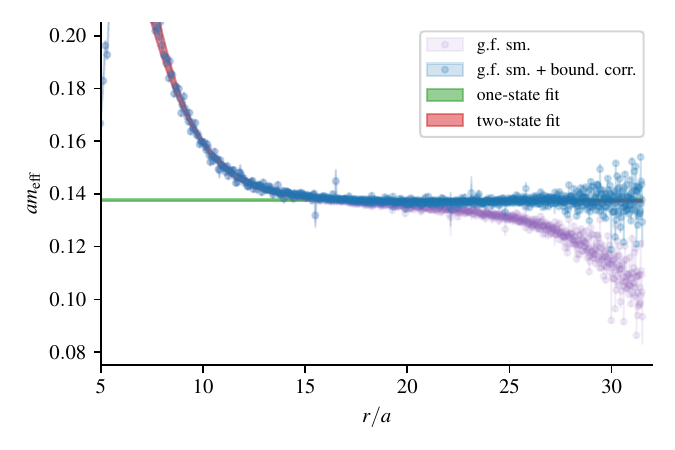}
\caption{Effective mass corresponding to the radial pion correlator $\mathring{C}_{PP}(r)$. The purple points are obtained applying the infinite-volume formula for the long-distance behavior in eq.~\eqref{eq:CPPdef}, whereas the blue points are obtained by accounting for finite-volume effects, eq.~\eqref{eq:FPP_final}. The green line shows the mass obtained from the best ``one-state'' fit to the correlator and the red curve shows the effective mass corresponding to the best ``two-state'' fit to the correlator as described in the main text.}\label{fig:pion_effmass}
\end{figure}

The simplest observable that we consider is the pion mass $m_\pi$, extracted from the position-space pseudoscalar correlator whose large-$|x|$ behavior is given in eq.~\eqref{eq:CPPdef}.
The rotational symmetry of the position-space correlator in the continuum and infinite volume is broken by the finite lattice spacing, with directions that are not equivalent under hypercubic symmetry contributing different cut-off effects.
In principle, the many different hypercubic-inequivalent directions provide us with considerable freedom in defining our hadronic observables;
however, to the purpose of this initial study, we limit ourselves to the correlator $\mathring{C}_{PP}(r)$ as function of the radial distance only as introduced in eq.~\eqref{eq:CPPring_def}.
This is obtained by averaging uniformly over hyperspheres of fixed $r$ in a way similar to the one discussed in section~\ref{sec:positionspace}. At finite lattice spacing, the integral over the angles $\int\dd\Omega_4$ is replaced by the sum
\begin{equation}
\label{eq:CPPring_lattice_def}
\mathring{C}_{PP}(r) = \frac{1}{\mathrm{r}_4(r^2/a^2)} \sum_{|x|=r} C_{PP}(x) ,
\end{equation}
where $\mathrm{r}_4(n)=8\sum_{d\mid n,4\nmid d} d$ is the number integer 4-vectors $z\in\mathbb{Z}^4$ satisfying $z^2=n$.\footnote{\url{https://oeis.org/A000118}}

Comparing this description of the lattice data with the long-$r$ behavior of eq.~\eqref{eq:CPPdef}, following eq.~\eqref{eq:numerical_effective_mass} we numerically solve for $m_\pi$ as a function of $r$.
It is worth noting that since there exists an $x$ such that $|x|=r$ for all integer values of $r^2/a^2$, the density in $r$ of the available lattice data increases with $r$.
We apply eq.~\eqref{eq:numerical_effective_mass} choosing $\Delta$ as the closest value to one lattice unit that is available, that is, such that $(r+\Delta)^2/a^2\in\mathbb{N}$, as we observe that this produces a much smoother effective mass than using, for instance, the smallest possible value of $\Delta$.

\begin{figure}[t]
\centering
\includegraphics{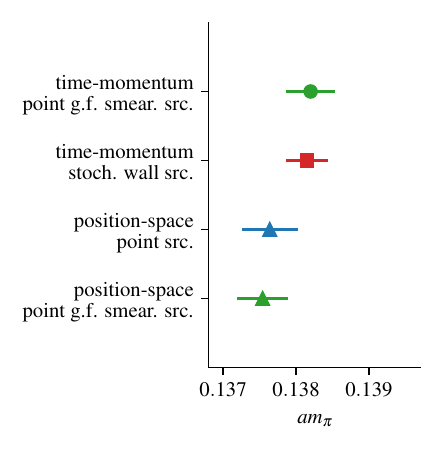}\includegraphics{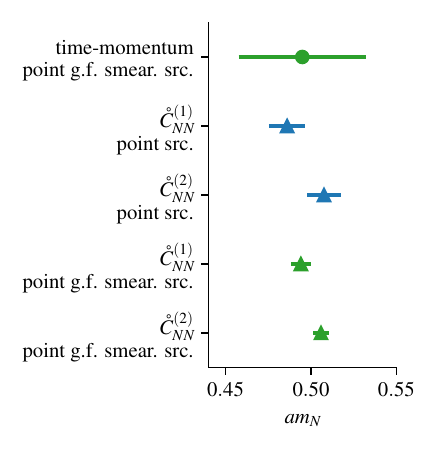}
\caption{\textbf{Left}: comparison of the value of $m_\pi$ obtained from the ``one-state'' fit for various choices of correlators and smearing. \textbf{Right}: comparison of the value of $m_N$ obtained from the ``one-state'' fit for various choices of correlators and smearing.}\label{fig:pion_comparison}\label{fig:nucleon_comparison}
\end{figure}

The result for the effective mass as a function of $r$ without correcting for boundary effects are shown in figure~\ref{fig:pion_effmass} in purple color, where it is evident that the determination is affected by significant boundary effects.
We therefore apply the correction described in section~\ref{sec:positionspace} and employ eq.~\eqref{eq:FPP_final} to solve for the effective mass against our lattice data using eq.~\eqref{eq:numerical_effective_mass}.
The resulting effective mass shown in blue in figure~\ref{fig:pion_effmass} is flat at large $r$, confirming that we are able to successfully account for boundary effects.
In the same plot we also show the results of a direct ``one-state'' fit to lattice data with the $m_\pi$ and $c_P$ parameters left free, restricted to the region in $r$ where there is an effective mass plateau, and of another ``two-state'' fit with an added ``excited state'' term $a_1 \frac{m_1}{r} K_1(m_1 r)$ with two extra parameters $a_1$, $m_1>m_\pi$.
In both cases, the fitted value of $m_\pi$ is compatible with the effective mass plateau value.

The results of the position-space determination of $m_\pi$ both with or without gradient-flow smearing are given in table~\ref{tab:point_sources_results_1} and shown in the left panel of figure~\ref{fig:pion_comparison}.
The result obtained without applying smearing to the sources is
\begin{equation}
\label{eq:pion_result}
m_\pi = \num{0.1376(4)}/a \approx \SI{288.9(8)}{\MeV} .
\end{equation}
We observe that, in the case of the pseudoscalar correlator, smearing does not lead to a better determination.
However, the position-space results have a better statistical precision than the results (also given in table~\ref{tab:point_sources_results_1}) obtained on the same point sources using the standard momentum-space techniques, including when $3d$ fermion smearing is used.

\subsubsection{Nucleon mass}
\label{sec:nucleon_mass}

\begin{figure}[t]
\centering
\includegraphics{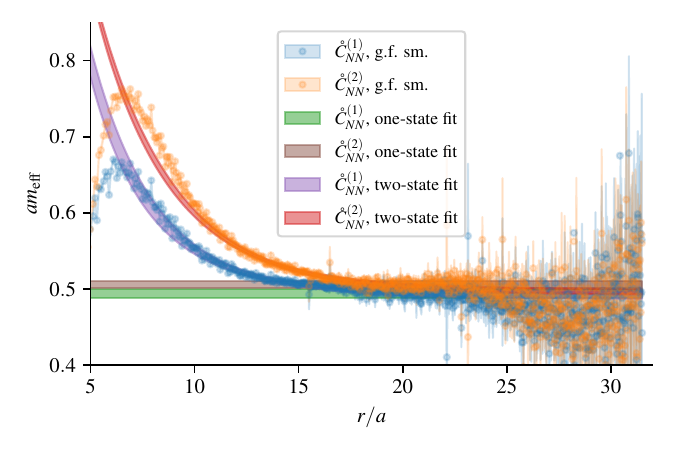}
\caption{Effective mass corresponding to the radial nucleon correlator $\mathring{C}_{NN}(r)$. The blue and orange points are obtained by applying the long-distance behavior to the $\tr\mathring{C}_{NN}$ and $\tr\slashed{x}\mathring{C}_{NN}$ contractions in eqs.~\eqref{eq:CNNring1_def} and \eqref{eq:CNNring2_def}, respectively. The horizontal lines show the mass obtained from the best ``one-state'' fit to the $\tr\mathring{C}_{NN}$ (green) and $\tr\slashed{x}\mathring{C}_{NN}$ (brown) correlators. The shaded curves show the effective mass corresponding to the best ``two-state'' fit to the $\tr\mathring{C}_{NN}$ (purple) and $\tr\slashed{x}\mathring{C}_{NN}$ (red) correlators described in the main text.}\label{fig:nucleon_effmass}
\end{figure}

The next hadronic observable that we considered is the nucleon mass $m_N$ extracted from the correlator of two $uud$ spinor fields in eq.~\eqref{eq:uud_spinor}, which in position space has the large-$|x|$ behavior given in eq.~\eqref{eq:CNNdef}. As in the case of the pseudoscalar correlator, in this work we consider only the correlator data averaged over all angles as an estimator of the radial correlator, which in this case takes the form of two contractions $\mathring{C}_{NN}^{(1)}(r)$ and $\mathring{C}_{NN}^{(2)}(r)$, defined as in eqs.~\eqref{eq:CNNring1_def} and~\eqref{eq:CNNring2_def} respectively with the $\int\dd\Omega_4\to\mathrm{r}_4^{-1}(r^2/a^2)\sum_{|x|=r}$ replacement as in eq.~\eqref{eq:CPPring_lattice_def}. We determine the effective masses from both correlators solving eq.~\eqref{eq:numerical_effective_mass}, again with $\Delta\approx 1a$, and taking into account that, in principle, the results of the two contractions can be different due to different discretization effects. The results are shown in figure~\ref{fig:nucleon_effmass} in blue and orange for $\mathring{C}_{NN}^{(1)}$ and $\mathring{C}_{NN}^{(2)}$ respectively, where we observe a boundary effect at large $r$ that leads to an increase of the effective mass. In contrast with the pseudoscalar correlator case, the main boundary contribution to the nucleon correlator does not come from the mirror images considered in section~\ref{sec:positionspace}, which fall off much faster than in the case of the pion, but from the propagation of intermediate $N\pi$ states.

We also perform a direct fit to the radial correlators using the appropriate large-$r$ description and with free $m_N$ and $c_N$ parameters, restricting the fit range in $r$ to the plateau region up to $r_{\mathrm{max}}=24a$ to avoid uncontrolled boundary effects. We note however that we expect any master-field calculation will have a sufficiently large volume that boundary effects become irrelevant. We also consider a fit that includes an extra factor of $[1 + a_1 \frac{m_\pi}{r} K_1(m_\pi r)]$, with a free amplitude parameter $a_1$ and a mass parameter fixed to $m_\pi$, and we observe that this model effectively describes the correlator data for both $\mathring{C}_{NN}^{(1)}$ and $\mathring{C}_{NN}^{(2)}$ to smaller values of $r$.

The results of the position-space determination of $m_N$ from both $\mathring{C}_{NN}^{(1)}$ and $\mathring{C}_{NN}^{(2)}$, with or without gradient-flow smearing, are given in table~\ref{tab:point_sources_results_1} and shown in the right panel of figure~\ref{fig:nucleon_comparison}. From $\mathring{C}_{NN}^{(1)}$ using gradient-flow smearing, we obtain the value
\begin{equation}
m_N = \num{0.494(6)}/a \approx \SI{1037(13)}{\MeV} .
\end{equation}
The nucleon mass extracted from $\mathring{C}_{NN}^{(2)}$ has a similar error but is systematically larger both with or without smearing, which is compatible with the fact that $\mathring{C}_{NN}^{(1)}$ and $\mathring{C}_{NN}^{(2)}$ can have different discretization effects. Using the same sources we obtain more precise result from position space compared with TMR; for discussion of the error scaling based on the Parisi-Lepage argument see appendix~\ref{app:parisi}. Contrary to the pion case, smearing has a visible impact on the determination of the nucleon mass. Applying $3d$-fermion smearing reduces the error on $m_N$ from the momentum-projected correlator by one third. Gradient-flow smearing also improves the already-smaller error of the position-space estimator by a factor of two. However, this comes at the cost of a distortion of the position-space correlator at short distances, within a range proportional to $\sqrt{t_{\mathrm{flow}}}$, that is visible in figure~\ref{fig:nucleon_effmass} at $r\lesssim 8a$. Including the extra factor of $[1 + a_1 \frac{m_\pi}{r} K_1(m_\pi r)]$ in the fit model yields compatible results with or without smearing as long as $t_{\mathrm{min}}\geq 8a$. Empirically, the fitted parameter $a_1$ is of the same order irrespective of smearing, suggesting that the excited states modelled by the extra factor are not suppressed by gradient-flow smearing. This is different from $3d$-fermion or other forms of smearing usually employed in the momentum-projected case, which generally contribute to suppressing the amplitude of excited states.

\subsubsection{Pion decay constant}

\begin{figure}[t]
\centering
\includegraphics{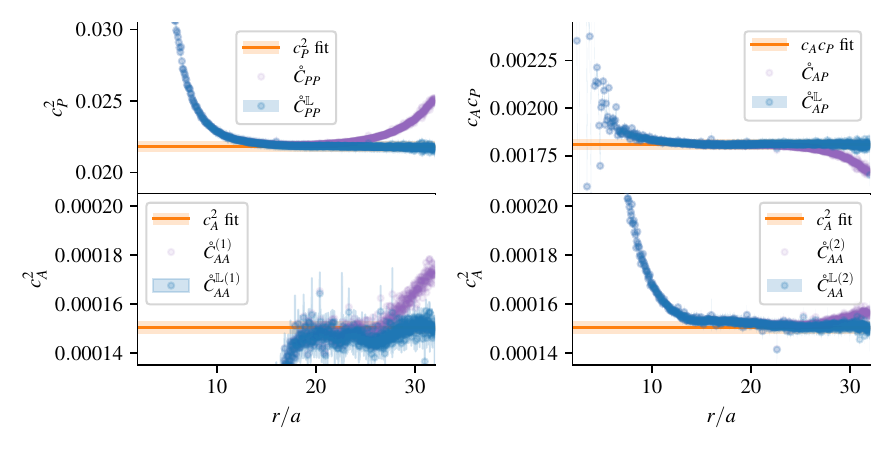}
\caption{Ratio of $\mathring{C}_{PP}$ (top left), $\mathring{C}_{AP}$ (top right), $\mathring{C}_{AA}^{(1)}$ (bottom right) and $\mathring{C}_{AA}^{(2)}$ (bottom left) correlator data to the fitted asymptotic behavior based on eqs.~\eqref{eq:Cring_asym}, \eqref{eq:CAAring1_def} and~\eqref{eq:CAAring2_def} respectively, with the fitted prefactors of $c_A$ and $c_P$ set to unity in the ratio. The purple points show the ratio obtained when omitting the boundary effects in the denominator. The blue points show instead the result when boundary effects are included, with the light blue error band representing the statistical error on the correlator data. The orange band shows the result of the fit for the normalization of each correlator, that determines the modulus of the $c_A$ and $c_P$ amplitude parameters and in turn the bare pion decay constant.}\label{fig:decay_constant}
\end{figure}

To compute the pion decay constant, we perform a combined fit of the pseudoscalar correlator in the radial direction $\mathring{C}_{PP}(r)$ defined on the lattice in eq.~\eqref{eq:CPPring_lattice_def}, together with the lattice version of $\mathring{C}_{AP}(r)$ in eq.~\eqref{eq:Cring_asym} and the two contractions $\mathring{C}_{AA}^{(1)}(r)$ and $\mathring{C}_{AA}^{(2)}(r)$ in eqs.~\eqref{eq:CAAring1_def} and~\eqref{eq:CAAring1_def},
\begin{align}
\mathring{C}_{AP}(r) &= \frac{1}{\mathrm{r}_4(r^2/a^2)} \sum_{|x|=r} x_\mu C_{AP,\mu}(x)
\sim \frac{c_Ac_P}{4\pi^2} K_2(m_\pi r) , \label{eq:CAPring_lattice_def} \\
\mathring{C}_{AA}^{(1)}(r) &= \frac{1}{\mathrm{r}_4(r^2/a^2)} \sum_{|x|=r} \delta_{\mu\nu} C_{AA,\mu\nu}(x)
\sim \frac{c_A^2}{4\pi^2} \frac{m_\pi}{r} K_1(m_\pi r) , \label{eq:CAAring1_lattice_def} \\
\mathring{C}_{AA}^{(2)}(r) &= \frac{1}{\mathrm{r}_4(r^2/a^2)} \sum_{|x|=r} x_\mu x_\nu C_{AA,\mu\nu}(x)
\sim \frac{c_A^2}{4\pi^2} \left[ m_\pi r K_1(m_\pi r) + 3 K_2(m_\pi r) \right] . \label{eq:CAAring2_lattice_def}
\end{align}
Using only the correlator data without any smearing of the local $A_\mu$ currents and $P$ densities, the combined fit allows us to simultaneously extract the mass $m_\pi$ and the amplitudes $c_P$ and $c_A$. As in the case of the pion mass discussed in section~\ref{sec:position_space_results_pion}, agreement with the data at the longest distances is obtained only if the boundary effects are accounted for in each of the four correlators in the combined fit model, which is obtained modifying the $r\to\infty$ asymptotic behavior of eqs.~\eqref{eq:CPPring_lattice_def}, \eqref{eq:CAPring_lattice_def}, \eqref{eq:CAAring1_lattice_def} and~\eqref{eq:CAAring2_lattice_def} according to the discussion in section~\ref{sec:positionspace}. At short distances, the data for $\mathring{C}_{AA}^{(1)}(r)$ has the opposite sign, and agrees with eq.~\eqref{eq:CAAring1_lattice_def} only at relatively large $r$. Thus, we fit $\mathring{C}_{AA}^{(1)}(r)$ data starting from $r_{\mathrm{min}}=24a$, and $\mathring{C}_{PP}(r)$, $\mathring{C}_{AP}(r)$ and $\mathring{C}_{AA}^{(2)}(r)$ starting from $19a$, $16a$ and $19a$ respectively. The results of the fit are
\begin{equation}
\label{eq:decay_contant_fit_results}
am_\pi = \num{0.1371(5)} , \qquad a^2c_P = \num{0.1477(13)} , \qquad a^2c_A = \num{0.01227(11)} .
\end{equation}
We observe that the pion mass obtained with this combined fit is smaller but compatible with the value given in eq.~\eqref{eq:pion_result}, obtained from the effective mass plateau of the $\mathring{C}_{PP}(r)$ correlator only and gradient-flow smearing.

Turning to the main focus of this section, these fit results yield a value of the decay constant
\begin{equation}
f_\pi^{\mathrm{bare}} = \frac{c_A}{m_\pi} = \num{0.0895(7)}/a \approx 187.8(1.6) \, \text{MeV}\,. %
\end{equation}
This position-space result for $f_\pi^{\mathrm{bare}}$ has a comparable statistical precision to the result obtained on the same point sources using the standard momentum space techniques, which are also given in table~\ref{tab:point_sources_results_2}.
The significant difference between the mean values of the highly-correlated position-space and zero-momentum results in the table can be attributed to the different discretization effects of the two methods.
A similarly significant difference is observed between the values of $f_\pi^{\mathrm{bare}}/m_\pi$ computed with position-space or zero-momentum methods.
In figure~\ref{fig:decay_constant} the lattice data for each of the four correlators are compared with the respective asymptotic behaviors both with and without the inclusion of boundary effects, reconstructed from the fit results in eq.~\eqref{eq:decay_contant_fit_results}.

\begin{table}[tb]
\centering
\begin{tabular}{lS[table-format=1.4(1)]S[table-format=1.3(1)]S[table-format=1.3(1)]S[table-format=1.3(1)]S[table-format=1.4(2)]}
\toprule
& {$af_\pi^{\mathrm{bare}}$} & {$f_\pi^{\mathrm{bare}}/m_\pi$} & \multicolumn{3}{c}{$\bar{\Pi}^{I=1}_{\mathrm{bare}}(-Q^2)$} \\
& & & {\SI{1}{\GeV\squared}} & {\SI{3}{\GeV\squared}} & {ratio} \\
\midrule
position-space & 0.0895(7) & 0.653(6) & 0.0568(4) & 0.0869(5) & 1.5288(28) \\
\midrule
zero-momentum & 0.0874(8) & 0.633(6) & 0.0584(7) & 0.0902(7) & 1.544(6) \\
\bottomrule
\end{tabular}
\caption{Pion decay constant, its ratio with the pion mass, and the HVP from position-space correlators on point sources, compared to zero-momentum projected correlator on the same point sources. The discrepancies between the two rows may be due to cutoff effects or, in the case of the HVP, due to failing to saturate the estimator in both cases.}\label{tab:point_sources_results_2}
\end{table}

\subsubsection{Hadronic vacuum polarization}

\begin{figure}[t]
\centering
\includegraphics{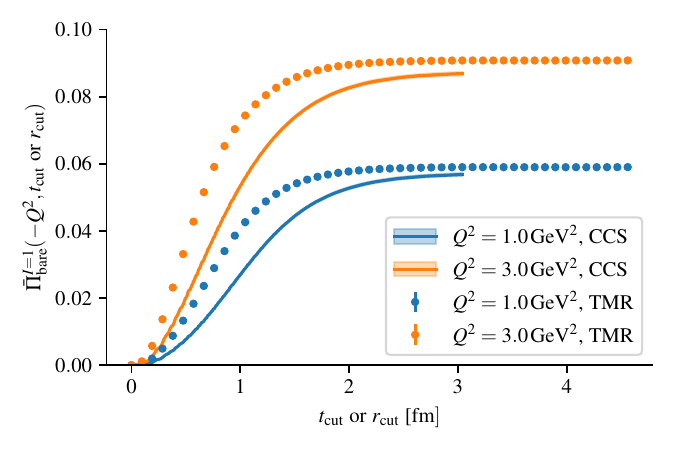}
\caption{Partial integral of $\bar{\Pi}^{I=1}(-Q^2)$ up to an Euclidean time $t_{\mathrm{cut}}$ (for the TMR method) or a radial distance $r_{\mathrm{cut}}$ (for the CCS method) at two values of $Q^2$, in different color. The statistical errors are smaller than the TMR point markers, and than the CCS line thickness.}\label{fig:PiHVP_comparison}
\end{figure}

Among the observables that can be extracted from the vector current, the hadronic vacuum polarization (HVP) contribution to the muon anomalous magnetic moment $(g-2)_\mu$ plays a prominent role. Due to the long-standing $(g-2)_\mu$ puzzle~\cite{Aoyama:2020ynm}, the HVP contribution is of major phenomenological interest and its determination at the sub-percent level is the objective of many large scale lattice efforts. Here, we limit ourselves to a computation of the related HVP function $\bar{\Pi}(-Q^2)$ defined in eq.~\eqref{eq:subtracted_HVP_TMR} at space-like momentum transfers of $Q^2=\SI{1}{\GeV\squared}$ and \SI{3}{\GeV\squared}; see also ref.~\cite{Ce:2022eix} for a much higher precision computation.

The covariant coordinate-space (CCS) method to extract $\bar{\Pi}(-Q^2)$ from position-space correlators has been introduced in refs.~\cite{Meyer:2017hjv,Ce:2018ziv} and amounts to integrating the sum of two contributions, each being the product of one of two contractions of the vector current, $\mathring{C}_{VV}^{(1)}(r)$ and $\mathring{C}_{VV}^{(2)}(r)$ from eq.~\eqref{eq:CVVring_def}, and its corresponding kernel, over the radial coordinate $r$, as shown in eq.~\eqref{eq:PiHVP_CCS}. In spirit, this is very similar to the traditional time-momentum representation (TMR) method where one computes the Euclidean-time integral of the product of the zero-momentum-projected vector correlator and a kernel. The choice of comparatively large $Q^2$ values helps in lowering the relative weight of the large $r$ distances in the kernel, which have large statistical uncertainties. In spite of this, we find that the extent in $r$ of the correlators that we have available is not enough to saturate the CCS integral, as shown in figure~\ref{fig:PiHVP_comparison} from the fact that the partial integral up to $r$ does not saturate at the largest $r$. This contrasts with the saturation of the Euclidean-time integral in the TMR method, also shown in figure~\ref{fig:PiHVP_comparison}, and confirms a major drawback of CCS methods already observed in ref.~\cite{Ce:2018ziv}: the need for very large volumes and long-range position-space correlators for the CCS integral to saturate. We also note that, even when the estimators are saturated, the resulting values are expected to differ by cutoff effects and finite-volume effects.

With this caveat, we perform a numerical test computing $\bar{\Pi}^{I=1}(-Q^2)$, the $I=1$ part of the HVP function. The computation of the HVP function relies on a genuine vector current, thus we use only the unsmeared meson operators.
We quote the results of $\bar{\Pi}^{I=1}(-Q^2)$ at two different values of $Q^2$,
\begin{equation}
  \label{eq:subtracted_Pi}
  \bar{\Pi}^{I=1}_{\mathrm{bare}}(-\SI{1}{\GeV\squared}) = \num{0.0568(4)} , \qquad \bar{\Pi}^{I=1}_{\mathrm{bare}}(-\SI{3}{\GeV\squared}) = \num{0.0869(5)} ,
\end{equation}
where the \emph{bare} subscript indicates that we are not including the renormalization factor $Z_V$ of the local vector current.
The error on $\bar{\Pi}^{I=1}_{\mathrm{bare}}$ is \SI{0.7}{\percent} and \SI{0.6}{\percent} at the smaller and larger value of $Q^2$, respectively.
This can be compared with the result obtained from the same correlator data using the TMR method: $\bar{\Pi}^{I=1}_{\mathrm{bare}}(-\SI{1}{\GeV\squared})=\num{0.0584(7)}$ and $\bar{\Pi}^{I=1}_{\mathrm{bare}}(-\SI{1}{\GeV\squared})=\num{0.0902(7)}$, with an error of \SI{1.2}{\percent} and \SI{0.8}{\percent} respectively.
These results are given in table~\ref{tab:point_sources_results_2}. While in both cases the statistical error of the traditional method is roughly \SI{50}{\percent} larger, this analysis does not take into account the systematics associated with the truncation of the integral, which is most likely larger in the CCS case.

To eliminate the need for the renormalization factor $Z_V$ of the local vector current, in table~\ref{tab:point_sources_results_2} we also quote the ratio between $\bar{\Pi}^{I=1}(-Q^2)$ at the two values of $Q^2$,
\begin{equation}
\label{eq:subtracted_Pi_ratio}
\frac{\bar{\Pi}^{I=1}(-\SI{3}{\GeV\squared})}{\bar{\Pi}^{I=1}(-\SI{1}{\GeV\squared})} = \num{1.5288(28)} .
\end{equation}
Taking the ratio also has an effect in reducing the relative statistical error to just \SI{0.18}{\percent}. This can be compared with the result obtained from the same correlator data using the TMR method where one finds \num{1.544(6)}.
For reference, in ref.~\cite{Ce:2022eix} the ratio in eq.~\eqref{eq:subtracted_Pi_ratio} has been computed to high precision using an $O(a)$-improved vector current on a CLS ensemble with a similar mass and $a\approx\SI{0.064}{\femto\metre}$ (N200), giving \num{1.5493(19)}, while in the continuum limit the same ratio evaluates to \num{1.484(14)}.

\section{Conclusions}
\label{sec:conclusions}

In this work, we have described and tested methods for using \emph{stochastic locality} to maximize the information extracted from a given set of numerical gauge fields. As is stressed throughout, the strategies used to optimally profit from locality must be examined on an observable-specific basis and should be considered both for the central values and the uncertainties of a given correlation function.

After providing standard correlator definitions and a specification of our lattice setup in section~\ref{sec:basics}, in section~\ref{sec:estimators} we have described our approach to estimate both central values and uncertainties. An important variation in our study is in the choice between correlators projected to definite spatial momentum, i.e.~in the time-momentum representation, and position-space correlators. We have introduced two classes of estimators for each of these: point sources and stochastic wall sources for the time-momentum representation, and point sources and stochastic grids for the position-space construction. We have additionally considered the effect of smearing in each case, so that our comparisons are more relevant to state-of-the art calculations that often employ the latter.

While our correlator estimators are well established in the literature, our approach for estimating the (co)variances is less explored. This is explained in sections~\ref{sec:variances} and \ref{sec:masterfield_error_saturation}. In a nutshell, the strategy is to view an estimator for the quantity of interest (e.g.~a two-point function with some fixed separation) as an observable with a footprint localized near some reference position $x$ (e.g.~the source position). One then subtracts the central value and defines a modified correlation function as the product of two instances of the subtracted observable with reference positions $x$ and $x+y$. A key consequence of stochastic locality is that, as the magnitude of $y$ is taken large, this function decays exponentially. The variance of the spatially averaged estimator of the target quantity is itself estimated from the integral over $y$ of the modified correlation function: we cut this off at long distance, leaving exponentially suppressed corrections.

In practice, the success of this method hinges on whether the integral value is saturated before finite-volume effects distort the estimator of the variance. In section \ref{sec:masterfield_error_saturation} we have demonstrated the feasibility of this across various observables. While the energy density at positive flow time, for example, has a variance that saturates very quickly, the situation is much more challenging for the zero-momentum pseudoscalar two-point function, especially at large source-sink separations. That the latter behaves poorly is intuitive: the quantum numbers, the spatial sum, and the source-sink separation all contribute to the observable having a large footprint, so spatially decorrelating a second instance of the quantity is challenging.

These results naturally led to the position-space approach presented in section~\ref{sec:positionspace}. While position-space correlation functions have more desirable locality properties, one expects excited-state and wrap-around effects to be more challenging without momentum projection. It is thus not obvious how the costs and benefits will balance in the final result. To explore this, we have compared the extraction of four different observables (the pion and nucleon mass, the pion decay constant, and the hadronic vacuum polarization function) using position-space and time-momentum representation correlators. Our results, summarized in section~\ref{sec:results}, show that position-space methods can give competitive determinations. While further investigation is needed, one generally finds lower statistical uncertainties from the position-space determinations, especially for the nucleon mass. In a full analysis, this may be offset by additional systematic uncertainties because our analysis of position-space correlators uses the approximation of continuum-like behavior at large distances, which is not required for analyzing momentum-projected correlators.

To summarize, we identify three take-home messages of this study: First, whenever multiple measurements are performed on a given gauge field, it is valuable to construct a modified correlator in order to study the spatial decorrelation of the observable. In the case of immediate decorrelation on the sampled set, one can treat the samples independently as with decorrelated gauge fields. Second, position-space correlation functions can be competitive in the determination of masses, matrix elements and other lattice observables based on two-point functions. Third, and finally, there is no need for a strict distinction between traditional calculations with some requisite number of configurations and master-field calculations with a requisite space-time volume. Instead, for any combined volume in the five-dimensional combination of space-time and Monte Carlo time, one can use data-driven studies of (auto)correlations to empirically examine the statistical precision for a given observable.

Future work in this vein includes the application of these strategies to larger space-time volumes on which the saturation of the variance correlator can be achieved for more observables. Additional work is also needed to rigorously treat the discretization and finite-volume effects in the fit functions used to extract observables from position-space quantities. Furthermore, while this work has considered the two extremes of standard momentum projection and position-space calculations, it would be instructive to consider a compromise between the two scenarios. In particular, momentum projection with a Gaussian wave packet or similar might well give an optimal balance between reducing the footprint and defining a useful estimation of the physical quantity of interest. An initial exploration of this is presented in appendix~\ref{app:truncated_sums}.

Despite the clear need for further exploration, it seems likely that careful use of stochastic locality, along the lines summarized in this article, will play a useful role in the next generation of lattice QCD calculations.

\begin{acknowledgments}
We warmly thank John Bulava and Martin L\"uscher for useful discussions and continued inspiration. This work made use of the openQCD software package \cite{Luscher:2012av,Francis:2019muy}.

MB thanks P.~Boyle, T.~Izubuchi and C.~Lehner for several useful discussions on the topic. The research of MB and MC is funded through the MUR program for young researchers ``Rita Levi Montalcini''.

MTH is supported by UKRI Future Leaders Fellowship MR/T019956/1, and in part by UK STFC grant ST/P000630/1.

The authors acknowledge the Texas Advanced Computing Center (TACC) at The University of Texas at Austin for providing HPC resources that have contributed to the research results reported within this paper. This research used resources of the National Energy Research Scientific Computing Center, which is supported by the Office of Science of the U.S. Department of Energy under Contract No. DE-AC02-05CH11231. Many simulations and measurements were also performed on a dedicated HPC cluster at CERN. The generous support of all these institutions is gratefully acknowledged. Furthermore we thank Kostas Orginos and Andr\'e Walker-Loud for their role supporting this effort and helping to make these resources available to us. Part of the analysis was performed using the pyobs library~\cite{PYOBS}.
\end{acknowledgments}

\appendix

\section{Extension of the Parisi-Lepage argument: Error volume scaling }
\label{app:parisi}

Decades ago, Parisi and Lepage explained the signal-to-noise problem in hadronic correlators~\cite{Parisi:1983ae, Lepage:1989hd}. Consider, for example, the zero-momentum nucleon correlator from a point source. Omitting the details of its color and spin structure, the essential aspect is that it involves three quark propagators:
\begin{equation}
\tilde C_{NN}(t) \sim \Re\int \dd^3\vec x
\left\langle S\left( (\vec x,t), (\vec 0,0) \right)^3 \right\rangle.
\end{equation}
Likewise, its variance has a contribution of the form
\begin{equation}
\operatorname{var}\left[ \tilde C_{NN}(t) \right]
\supset \int \dd^3\vec x \int \dd^3\vec y
\left\langle S\left( (\vec x,t), (\vec 0,0) \right)^3
S^*\left( (\vec y,t), (\vec 0,0) \right)^3
\right\rangle,
\end{equation}
with three quark propagators and three antiquark propagators. This can be understood as a correlation function in a partially quenched theory with six degenerate valence quarks, where the source interpolating operator is local, creating three quarks and three antiquarks, and the sink interpolating operator is bilocal, annihilating three quarks at one site and three antiquarks at the other. In general, the lightest state that can couple to these operators has three pion-like pseudoscalar mesons. At large $t$, the variance decays as $\exp(-E_0 t)$, where $E_0$ is equal to $3m_\pi$ up to corrections due to interactions among the mesons that fall off with a power of $1/L$. This leads to the well-known result that the signal-to-noise ratio of the nucleon correlator decays at large $t$ as $\exp(-(m_N-3m_\pi/2)t)$~\cite{Lepage:1989hd}.

The contribution from a single three-meson state to the variance correlator will fall off as $1/L^6$, which is compensated by the density of states. Therefore, for a better description one should take into account multiple states. In infinite volume, the continuum of three-meson states leads to a modified asymptotic behavior of the variance correlator, $\exp(-3m_\pi t)/t^3$, with faster decay. Therefore, at fixed $t$ we expect the variance to decrease with increasing $L$ as we transition from the regime where a single state dominates the variance to the regime where a continuum of states is relevant. One can model this by making several approximations: neglecting interactions among the mesons, assuming that only contributions from $\vec x \approx \vec y$ are relevant, and assuming that the couplings of all three-meson states to the interpolating operators are the same. We arrive at the three-meson contribution to the variance being proportional to
\begin{equation}
f_\text{var}(t,L) = \sum_{\vec p_1,\vec p_2} \frac{e^{-(E_{\vec p_1}+E_{\vec p_2}+E_{-(\vec p_1+\vec p_2)})t}}{8 E_{\vec p_1} E_{\vec p_2} E_{-(\vec p_1+\vec p_2)} L^6},
\end{equation}
where $E_{\vec p}=\sqrt{m_\pi^2+p^2}$ and the sums are over allowed finite-volume momenta. Note that in this simple model we have not accounted for the $J^P=0^+$ or $1^+$ quantum numbers to which the variance correlator is expected to couple.

\begin{figure}
\centering
\includegraphics{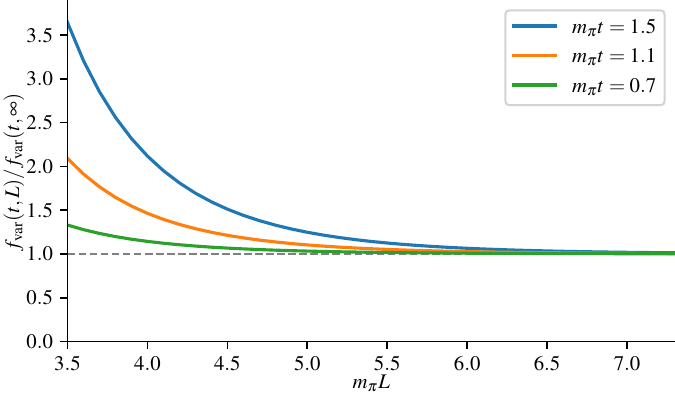}
\caption{Estimate of the three-meson contribution to the variance of the zero-momentum nucleon correlator versus box size, normalized to its value in infinite volume. Curves are shown for three values of $m_\pi t$: 0.7 and 1.5 correspond to $t\approx \SI{1}{\femto\metre}$ for $m_\pi=135$ and 290~MeV, respectively.}\label{fig:fvar_vs_mL}
\end{figure}

The dependence of this function on $L$ is shown in figure~\ref{fig:fvar_vs_mL}. At small values of $m_\pi L$, it is significantly above its infinite-volume value, and for larger values of $m_\pi t$, larger values of $m_\pi L$ are required to be close to infinite volume. It is also worth noting that if one assumes the computational cost scales with $L^3$, then for small values of $L$ this cost can be compensated by the decrease of variance with $L$: for $m_\pi t=1.5$, the point where the slope of $L^3 f_\text{var}(t,L)$ vanishes is $m_\pi L\approx 4$.

\begin{figure}
\centering
\includegraphics{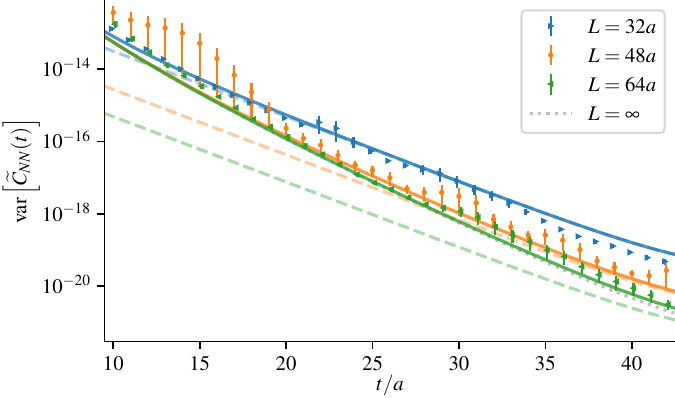}
\caption{Variance of zero-momentum nucleon correlator versus source-sink separation, for three lattice volumes and fixed pion mass: ensembles 32B, 48B, and 64B. Curves show the estimate $f_\text{var}$, multiplied by a single overall normalization factor to line up with the lattice data and with a modification $e^{-Et}\to e^{-Et}+e^{-E(L_t-t)}$ to model the effect of the finite time extent. The dotted curve shows the estimate in infinite spatial volume and the dashed curves show the contribution to $f_\text{var}$ from three mesons at rest.}\label{fig:var_vs_t}
\end{figure}

Our estimate is confronted with data for three different lattice volumes with pion mass 293~MeV in figure~\ref{fig:var_vs_t}. At large $t$, the lowest-lying three-meson state is the dominant contribution to the variance, and the scaling of this state's amplitude as $L^{-6}$ produces a smaller variance in the larger volumes. At smaller $t$, higher-lying three-meson states contribute more in larger volumes, reducing the volume dependence of the variance.

For the radially averaged position-space nucleon correlator, the same approximations lead to a variance that decays proportionally to $[C_{PP}(r)/r]^3$ at large distance $r$. Thus, asymptotically the nucleon signal-to-noise ratio behaves as follows:
\begin{enumerate}
\item zero momentum (finite volume): $e^{-(m_N-3m_\pi/2)t}$,
\item zero momentum (infinite volume): $t^{1.5}e^{-(m_N-3m_\pi/2)t}$,
\item radially averaged: $r^{2.25}e^{-(m_N-3m_\pi/2)r}$.
\end{enumerate}
This somewhat improved asymptotic behavior may partially explain the reduced uncertainty we obtained for the nucleon mass in section~\ref{sec:nucleon_mass}.

\section{Truncating three-dimensional sums for correlator estimators in large volumes}
\label{app:truncated_sums}

A generic strategy for reducing the noise in correlation functions that involve a sum over some of the lattice dimensions is to approximate the sum by restricting it to a sub-volume. If one can include the dominant contributions to the signal while excluding noisy long-distance contributions, this offers the prospect of trading a small bias for a reduction in statistical uncertainty. This idea has been explored in ref.~\cite{Liu:2017man} but is also widely used e.g.\ when restricting temporal sums in calculations of the hadronic vacuum polarization contribution to the muon anomalous magnetic moment.

\begin{figure}
\centering
\includegraphics[width=0.495\textwidth]{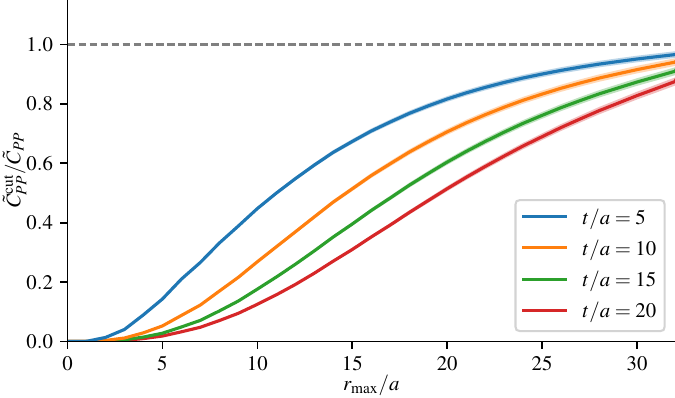}
\includegraphics[width=0.495\textwidth]{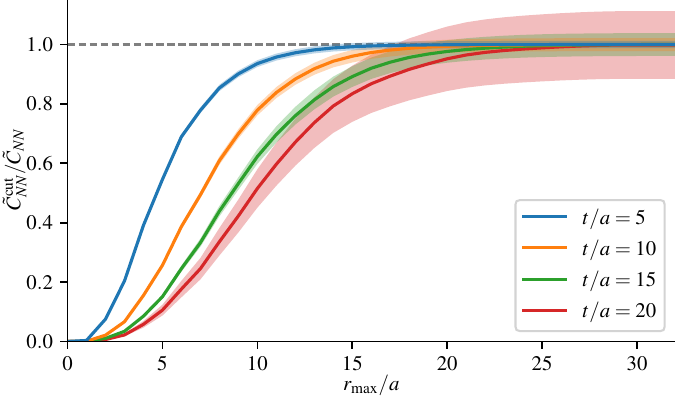}
\caption{Truncated zero-momentum pion (left) and nucleon (right) correlators $\tilde C^\text{cut}(t,r_\text{max})$ versus $r_\text{max}$, normalized by the central value of $\tilde C(t)$. Data are shown for four values of $t$.}\label{fig:corr_cut}
\end{figure}

To this end, we define the truncated zero-momentum correlator
\begin{equation}
\begin{aligned}
\tilde C^\text{cut}(t,r_\text{max}) &\equiv \int \dd^3\vec x\, \theta(r_\text{max}-|\vec x|) C(\vec x,t)
\xrightarrow{r_\text{max}\to\infty} \tilde C(t).
\end{aligned}
\end{equation}
Pion and nucleon correlators with gradient-flow smearing on the $64^3$ ensemble are shown in figure~\ref{fig:corr_cut}. In general, to reach the same fraction $\tilde C^\text{cut}/\tilde C$, larger source-sink separations $t$ require larger cutoff radii $r_\text{max}$. The pion correlator is slow to saturate: even $r_\text{max}=L/2$, i.e.\ summing over the largest ball that fits inside the lattice volume, is not consistent with the full sum. For the nucleon, the signal does saturate by $r_\text{max}=20a$. However, the noise also saturates, so that there is no statistical benefit from truncating the sum. This finding is consistent with the result of ref.~\cite{Liu:2017man} for connected diagrams: both the signal and the noise of the position-space nucleon correlator decay exponentially at large distance.

\section{Estimating the error of the error and other biases}
\label{app:errors}

First we examine the bias in our estimator $\llangle \Gamma_{\alpha\beta} \rrangle$ by considering
\begin{equation}
\begin{split}
\bigl\langle \llangle \Gamma_{\alpha\beta}(x) \rrangle \bigr\rangle = &
\frac{1}{N} \sum_z \langle \big[\obs_\alpha(x+z) -
\llangle \obs_\alpha \rrangle \big]
\big[ \obs_\beta(z) - \llangle \obs_\beta \rrangle \big] \rangle \,. \\
\end{split}
\end{equation}
Noting that
\begin{equation}
\frac{1}{N} \sum_z \langle \llangle \obs_\alpha \rrangle
\big[\obs_\beta(z) - \llangle \obs_\beta \rrangle \big] \rangle = 0 \,,
\end{equation}
for the remaining terms we obtain
\begin{equation}
\frac{1}{N} \sum_z \langle \obs_{\alpha}(x+z) \big[\obs_\beta(z) -
\llangle \obs_\beta \rrangle \big] \rangle =
\langle \obs_\alpha(x) \obs_\beta(0) \rangle -
\frac{1}{N^2} \sum_{z,z'} \langle \obs_\alpha(x+z) \obs_\beta(z') \rangle \,.
\end{equation}
By adding and subtracting $\langle \obs_\alpha \rangle \langle \obs_\beta \rangle$ we get the straightforward extension of the corresponding result in ref.~\cite{Wolff:2003sm}
\begin{equation}
\bigl\langle \llangle \Gamma_{\alpha\beta}(x) \rrangle \bigr\rangle =
\Gamma_{\alpha\beta}(x) - \frac{1}{N} C_{\alpha\beta} \,.
\label{eq:Gamma_bias}
\end{equation}

The second bias that we consider originates from the truncation of the integral over $\Gamma_{\alpha\beta}$. In eq.~(\ref{eq:Gamma}) periodic boundary conditions in all directions have been assumed, implying that we can average $N=V/a^4$ estimators of $\Gamma_{\alpha\beta}(x-y)$ (for all distances $x-y$). In such cases only an exponentially small bias is introduced. The situation is different if instead we consider the case with open or Dirichlet boundary conditions along at least one dimension. This situation is realized when the observables are known in a sub-domain of the lattice or if open boundaries are employed along the time direction for example. Taking this second simpler case for a lattice geometry of $L^3 \times T$, with $T$ denoting the temporal extent, we have
\begin{equation}
\frac{1}{N^2} \sum_{x,y} \Gamma_{\alpha\beta}(x-y) =
\frac{1}{N^2} \sum_{z_0, \vec z} \Gamma_{\alpha\beta}(z)
L^3 (T-z_0) \simeq \frac{1}{N} C_{\alpha\beta}
\Big(1 + O(1/(mT)) \Big) \,,
\end{equation}
with $m$ the mass scale dominating the behavior of $\Gamma_{\alpha\beta}$ at large distances. A similar correction is present in Markov chains \cite{Madras:1988ei}.

We now analyze the statistical error of the covariance matrix $\Cab{\alpha\beta}{(R)}$. To simplify the notation, we denote sums $\sum_{|x|\leq R}$ with $\widetilde \sum_{x}$ and we examine the covariance of $\llangle C_{\alpha\beta}(R) \rrangle$ with $\llangle C_{\gamma\delta}(R) \rrangle$:
\begin{equation}
\begin{aligned}
\operatorname{cov}\left( \llangle C_{\alpha\beta}(R) \rrangle,
\llangle C_{\gamma\delta}(R) \rrangle \right)
&= \widetilde\sum_{x,y} \left\langle
\left(\llangle \Gamma(x) \rrangle - \Gamma'(x)\right)_{\alpha\beta}
\left(\llangle \Gamma(y) \rrangle - \Gamma'(y)\right)_{\gamma\delta}
\right\rangle \\
&= \widetilde \sum_{x,y} \frac{1}{N^2} \sum_{z,z'} \left\langle
\left[\delta \obs_\alpha(x+z) \delta \obs_\beta(z) \right]
\left[ \delta \obs_\gamma(y+z^\prime) \delta \obs_\delta(z^\prime) \right]
\right\rangle \\
&\quad - \widetilde \sum_{x,y} \Gamma'_{\alpha\beta}(x) \Gamma'_{\gamma\delta}(y),
\end{aligned}
\end{equation}
where $\Gamma'_{\alpha\beta}(x)\equiv \langle \llangle
\Gamma_{\alpha\beta}(x)\rrangle \rangle$ is the biased expectation
value of our estimator $\llangle\Gamma_{\alpha\beta}\rrangle$.

Setting $\alpha=\gamma$ and $\beta=\delta$ returns precisely $\operatorname{var}(\Cab{\alpha\beta}{(R)})$. Focusing on the four-point function, we neglect the fully connected part, relevant only when $x,y,z,z'$ are all close to each other, and like in ref.~\cite{Wolff:2003sm} consider solely the factorized expectation values\footnote{We performed numerical checks and verified that this approximation works well in practice.}
\begin{multline}
\frac{1}{N^2} \sum_{z,z'}
\langle \big[\delta \obs_\alpha(x+z) \delta \obs_\beta(z) \big]
\big[ \delta \obs_\gamma(y+z^\prime) \delta \obs_\delta(z^\prime) \big] \rangle
\approx \Gamma'_{\alpha\beta}(x) \Gamma'_{\delta\gamma}(y) \\ + \frac{1}{N^2}
\sum_{z,z'} \Big[
\Gamma_{\alpha\gamma}(x+z-y-z')\Gamma_{\beta\delta}(z-z') +
\Gamma_{\alpha\delta}(x+z-z') \Gamma_{\beta\gamma}(z-y-z')
\Big] \,.
\end{multline}
In the second line we have explicitly neglected the bias correction derived in eq.~\eqref{eq:Gamma_bias}, since it amounts to an effect of higher order in $1/N$, while we kept it in the first term to cancel the corresponding one in $\operatorname{cov}\left( \llangle C_{\alpha\beta}(R) \rrangle, \llangle C_{\gamma\delta}(R) \rrangle \right)$.
Noting that the equation above depends only on the difference $z-z'$, we
introduce $t\to z-z'$ and $t\to z-z'-y$ in the last two terms respectively,
simplify the sum over $z'$, thus obtaining
\begin{equation}
\widetilde \sum_{x,y} \frac{1}{N} \sum_{t} \Big[
\Gamma_{\alpha\gamma}(x-y+t) \Gamma_{\beta\delta}(t) +
\Gamma_{\alpha\delta}(x+y+t) \Gamma_{\beta\gamma}(t) \Big] \,.
\end{equation}
Thanks to the rapid fall-off of $\Gamma_{\beta\delta}(t)$ and $\Gamma_{\beta\gamma}(t)$ at large separations, we can restrict the sum over $t$ to $|t|\leq R$ up to exponentially-suppressed corrections.
Similarly, we can substitute $x \mp y+t\to x$ in the first and second terms respectively with no changes to the support of $x$ up to exponentially-suppressed corrections.
This leads us to
\begin{equation}
\operatorname{cov}\left( \llangle C_{\alpha\beta}(R)
\rrangle, \llangle C_{\gamma\delta}(R) \rrangle \right) \approx
\frac{N(R)}{N} \big[
C_{\alpha\gamma} \, C_{\beta\delta} +
C_{\alpha\delta} \, C_{\beta\gamma} \big] \,,
\label{e:error_of_error}
\end{equation}
where $N(R) = \widetilde\sum_y = \sum_y\theta(R - |y|)$.

\section{An automatic windowing procedure for the determination of the master-field error}
\label{app:window}

In this appendix we focus on the errors, namely $C_{\alpha\alpha}(R)$, and we drop the corresponding subscripts (also from $\Gamma_{\alpha\alpha}$). Starting from the following Ansatz for the asymptotic behavior of the correlation function $\Gamma$, with integer $k \geq 0$,
\begin{equation}
\Gamma(x) \overset{x>0}{\sim} \frac{\Gamma(0)}{|x|^k} e^{-m |x|} \,,
\end{equation}
and assuming to use our error estimators in $D$ dimensions,
the relative systematic error due to the truncation of the sum in eq.~\eqref{eq:Cbar} is approximated by
\begin{equation}
\frac{\delta_\mathrm{sys} C(R)}{C} =
\frac{\tau-\tau(R)}{\tau} =
\frac{\Omega_D}{\tau}
\int_{R}^\infty \dd r \, r^{D-1-k} \, e^{-m r}=
\frac{\Omega_D}{\tau} \frac{\Gamma(D-k, mR)}{m^{D-k}} \,.
\label{eq:tau_model}
\end{equation}
In the equation above, $\tau$ represents the integrated correlation volume, $\Gamma(s,b)$ the upper incomplete gamma function, defined as
\begin{equation}
\Gamma(s,b) = \int_b^\infty \dd z \, z^{s-1} \, e^{-z}\,,
\label{eq:gammainc}
\end{equation}
and $\Omega_D=2\pi^{D/2}/\Gamma(D/2)$ is the complete solid angle in $D$ dimensions.

A good balance with the error of the error derived in eq.~\eqref{e:error_of_error} is achieved for the minimal value of $R$ where the derivative
\begin{equation}
g(R) = \frac{\partial}{\partial R}
\bigg[\frac{\delta_\mathrm{sys} C(R) + \sqrt{\operatorname{var}\left( \llangle C(R) \rrangle \right)}}{C}
\bigg] =
\frac{e^{-mR} R^{D-1-k} \Omega_D}{\tau} +
\sqrt{\frac{D R^{D-2} \Omega_D}{2 N}}
\end{equation}
becomes positive. Note that above we used eq.~\eqref{e:error_of_error} with the continuum version of $N(R)$. Following Ulli Wolff's suggestion in ref.~\cite{Wolff:2003sm}, we replace $\tau$ with the calculated $\tau(R) \leq \tau$ and $m$ with the effective mass (from eq.~\eqref{eq:tau_model})
\begin{equation}
m(R)^{D-k} \equiv \frac{\Omega_D \, \Gamma(D-k, 0)} {\tau(R)} \,,
\end{equation}
thus obtaining
\begin{equation}
g(R) = \Omega_D \frac{e^{-S_\tau m(R)R} R^{D-1-k} }{\tau(R)}
+ \sqrt{\frac{D \Omega_D}{2 V}} R^{D/2-1} \,.
\end{equation}
We introduced the parameter $S_\tau$ which plays a similar role as in ref.~\cite{Wolff:2003sm}, while the parameter $k$ is a new tunable parameter. From our experiments $S_\tau \simeq 2$ and $k=0$ return acceptable windows. This procedure is implemented in the pyobs library~\cite{PYOBS}.

\section{Long-\texorpdfstring{$T$}{T} simulations, a master-field variation using very cold lattices}
\label{sec:app_longT}

The long-$T$ approach was put forward in ref.~\cite{Bruno:2022ljo} as a variation of the master-field idea. To summarize, the master-field approach presents a strategy to manage contaminations originating from the topology freezing in simulations, aside from proposing a new statistical method for evaluating observables and their uncertainties as is studied in the main body of this work.

To this end, note that the effect of freezing scales as $1/V$ for sufficiently large space-time volume $V$, while statistical uncertainties are only suppressed by $1/\sqrt{V}$. For a fixed number of gauge fields they are therefore guaranteed to dominate over freezing effects as the space-time volume increases. An effective suppression of topological effects can therefore be reached by increasing all four dimensions but also by increasing only one dimension while keeping the others fixed. Choosing to elongate the $T$ dimension while keeping the spatial volume fixed at a ``traditional'' size, e.g.\ at the small volume of $L=32a$ in our simulations here, we arrive at the long-$T$ approach.

In our previous study, we were in particular interested in the suppression of topological freezing with $T$, dubbed defrosting, in simulations with a lattice spacing where topology freezing becomes a problem, i.e.\ for $a$ approximately $0.055$~fm for our choice of action. However, in that study, resource limitations put a comparison of results in the large space-time volumes achieved by the long-$T$ method and corresponding spatial volumes out of reach. Furthermore, due to the onset of topology freezing, including small volumes in a comparison of topological observables became more difficult.
The situation in our case here is different: With the selected lattice spacing of $a=0.094~$fm, topology freezing is not problematic, and we are able to generate and compare lattices in a consistent manner.

\begin{figure}
\centering
\includegraphics{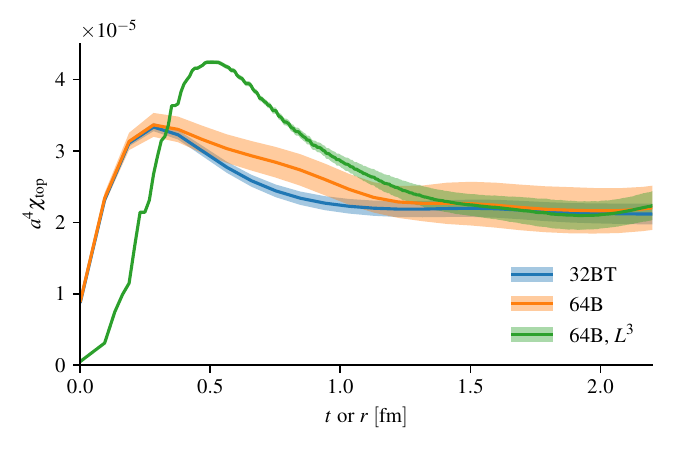}
\caption{Truncated sum for topological susceptibility versus Euclidean time $t$ [eq.~\eqref{eq:topological_susceptibility_integral_def}] or three-dimensional spatial distance $r$ with label ``$L^3$'' [eq.~\eqref{eq:topological_susceptibility_integral_def_2}]; the asymptotic plateau values yield $\chi_\text{top}$. The two ensembles examined here differ only in geometry while having the same four-volume: the 32BT follows the long-$T$ approach and has an elongated direction $T=768a$ with fixed ``traditional'' spatial size $L=32a$, whereas the 64B has an increased spatial size $L=64a$ with $T=96a$. Eq.~\eqref{eq:topological_susceptibility_integral_def} produces a larger statistical error for ensemble 64B than for 32BT because its spatial sum includes more noisy contributions from large spatial separations of the topological charge density; the smaller footprint of the estimator in eq.~\eqref{eq:topological_susceptibility_integral_def_2} produces a smaller error approaching that of ensemble 32BT.}\label{fig:longT}
\end{figure}

Our main observable is the topological susceptibility $\chi_\mathrm{top}$. We estimate it from the truncated sum (see e.g.\ ref~\cite{Bruno:2014ova})
\begin{equation}
\llangle \chi_\mathrm{top}(t) \rrangle = \frac{a}{T} \sum_{x_0} \chi_\mathrm{top}(t; x_0) \,,
\label{eq:topological_susceptibility_average}
\end{equation}
with
\begin{equation}
\label{eq:topological_susceptibility_integral_def}
\chi_{\mathrm{top}}(t ; x_0) = \frac{a^7}{L^3} \sum_{-t\leq y_0\leq t} \sum_{\vec x,\vec y \in L^3} q(\vec y,x_0+y_0) q(\vec x, x_0) \,, \quad
\langle \chi_\mathrm{top}(t;x_0) \rangle \overset{t \gg 0}{\approx} \chi_\mathrm{top} \,,
\end{equation}
and
\begin{equation}
q(x) = - \frac{1}{32\pi^2} \epsilon_{\mu\nu\rho\sigma} {\rm Tr}[ F_{\mu\nu}(x)F_{\rho\sigma}(x)]~.
\end{equation}
$F_{\mu\nu}(x)$ represents the field-strength tensor which we determine at positive gradient-flow time \cite{Luscher:2010iy}, specifically at $t_{\mathrm{flow}}/ a^2 =3.125 $ equivalent to $\sqrt{8t_{\mathrm{flow}}} \simeq 0.47$~fm, using the clover discretization. In this appendix we examine two ensembles with equal space-time volumes and same bare parameters, namely the 32BT and 64B lattices (cf. table~\ref{tab:lat_ens}), and we use for both the same number of measurements (equally spaced in MD units). Therefore, we expect similar statistical uncertainties, which we estimate using correlations along Euclidean and Monte Carlo time for the 32BT and 64B ensemble respectively\footnote{For the long-T ensemble we follow the discussion in section~\ref{sec:variances}, namely we set $\mathcal O_{\alpha=t}(x_0) = \chi_\mathrm{top}(t ; x_0)$ and use $\Lambda_T$ from eq.~\eqref{eq:var_T_TL_LLL}. For the 64B we use the Monte Carlo fluctuations of $\llangle \chi_\mathrm{top}(t) \rrangle$ to estimate the error.}.
However, as we can see from the results in figure~\ref{fig:longT}, the 64B ensemble has in general larger errors. This effect is understood as originating from the zero-momentum projection, which for the 64B implies including pairs of points further apart (w.r.t.\ 32BT) which contribute only to the noise of the observable. To clarify this issue, on the 64B lattice, we estimate the topological susceptibility using a truncated spatial sum of the temporally-summed correlator,
\begin{equation}
\label{eq:topological_susceptibility_integral_def_2}
\chi_{\mathrm{top}}(r; \vec{x}) = \frac{a^5}{T } \sum_{x_0, y_0} \sum_{|\vec y|\leq r} q(\vec x + \vec y,y_0) q(\vec x, x_0) \,, \quad
\langle \chi_{\mathrm{top}}(r; \vec{x}) \rangle \overset{r \gg 0}{\approx} \chi_\mathrm{top} \,.
\end{equation}
The corresponding results are labelled in figure~\ref{fig:longT} as ``64B, $L^3$''. Their statistical error is closer to the 32BT zero-momentum estimator. A nice agreement of the asymptotic plateau values for all estimators of $\chi_{\mathrm{top}}$ is found, showing the effectiveness of the long-$T$ approach as a variation of the more general master-field paradigm.

\section{Observing and diagnosing ``exceptional'' configurations}
\label{app:exceptional}

\begin{figure}
\centering
\includegraphics{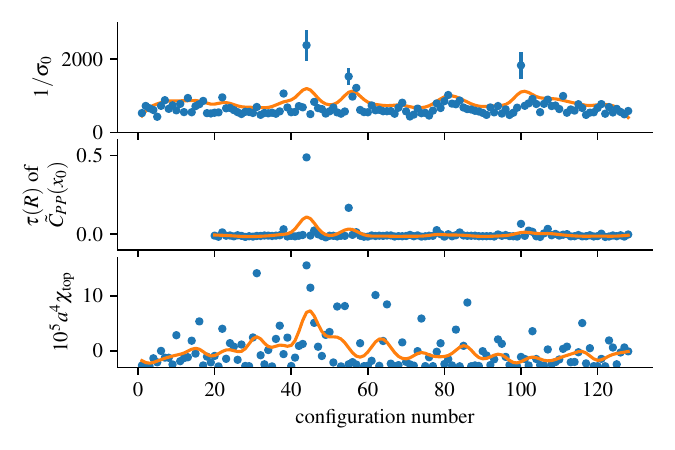}
\caption{Monte Carlo history plots of three quantities on the extended 64B ensemble. In the top row, we graph the inverse of the smallest singular value $\sigma_0$ of the Dirac operator with light quark mass. The central row shows the integrated correlation volume $\tau(R)$ of the pseudoscalar correlator $\tilde C_{PP}(x_0)$ as studied in section~\ref{sec:masterfield_error_saturation}, evaluated at $R=10a$ and $x_0=8a$. In the bottom row, we plot the Monte Carlo history of the topological charge susceptibility at flow-time $t_{\mathrm{flow}}/a^2=3.125$ (see also appendix~\ref{sec:app_longT}). For both observables in the central and bottom line, the ensemble average is subtracted and only the fluctuations around the mean are shown. The orange lines are MC histories smeared with a Gaussian filter with a width of three configurations.}\label{fig:spikes}
\end{figure}

While performing measurements on the 64B ensemble listed in table~\ref{tab:lat_ens}, we observed significant fluctuations in the Monte Carlo history of two observables: the topological susceptibility and the variance of the pseudoscalar correlator. These fluctuations resulted in spikes that showed an impact on the observable. For instance, when measured on all the available configurations, the topological susceptibility on 64B as shown in figure~\ref{fig:longT} was significantly different than the value on the ensemble 32BT, which has the same four-dimensional volume. Moreover, the error on the hadronic quantities measured in section~\ref{sec:results} was larger than expected.

To investigate the issue, we studied the spectrum of the light quark Dirac operator $D$ by computing $\sigma_0$, the square root of the smallest eigenvalue of the Hermitian operator $D^\dagger D$ using a subspace iteration algorithm. The Monte Carlo history of the two observables and of $1/\sigma_0$ is plotted in the three panels of figure~\ref{fig:spikes}.

In the central panel of figure~\ref{fig:spikes} we plot the fluctuations of the integrated correlation volume $\tau(R=10a)$, defined in eq.~\eqref{eq:tauD}, of the zero-momentum-projected pseudoscalar correlator $\tilde{C}_{PP}(x_0)$ at $x_0=8a$. Given the absence of a signal-to-noise problem, and an approximate behavior with four inverse powers of $D$, it is not surprising to see spikes in its Monte Carlo history coinciding with $1/\sigma_0$. This shows how the variance defined from stochastic locality may be used to monitor the presence of exceptional configurations.

In the bottom panel of figure~\ref{fig:spikes} we plot the fluctuations of the topological susceptibility defined using the global topological charge, which is equivalent to eq.~\eqref{eq:topological_susceptibility_average} with $t=T/2$, as measured on each configuration around the ensemble mean. The correlation of this gluonic observable with the low-lying spectrum of the Dirac operator is weaker and to highlight the presence of larger fluctuations in the first part of the Monte Carlo chain, we smear the Monte Carlo histories with a Gaussian kernel with width equal to 3 (in units of configuration numbers), represented by solid lines in figure~\ref{fig:spikes}.

Empirically, if we use only the second part of the Monte Carlo chain, avoiding the main spikes visible in figure~\ref{fig:spikes}, we observe that the topological susceptibility agrees between 64B and 32BT as shown in figure~\ref{fig:longT} and the error on hadronic quantities is reduced or stays approximately constant in spite of the lower statistics. Owing to these observations, we decided to limit all measurements used in the paper for this ensemble to the last 50 configurations (as reported in table~\ref{tab:lat_ens}), from 79 up to 128, of the Monte Carlo history in figure~\ref{fig:spikes}.

\bibliographystyle{JHEP}
\bibliography{refs.bib}

\end{document}